\documentclass{aa}  

\usepackage{graphicx}
\usepackage{txfonts}
\usepackage{xcolor}
\usepackage{subcaption}
\usepackage[colorlinks=true, allcolors=blue]{hyperref}

\begin{document}

    \title{Binary imposters: Mergers in massive hierarchical triple stars}
   
   \author{F. Kummer
          \inst{1}
          \and
          G. Simion\inst{1}
          \and
          S. Toonen\inst{1}
          \and A. de Koter\inst{1,2} 
          }

   \institute{Anton Pannekoek Institute for Astronomy, University of Amsterdam, Science Park 904, 1098 XH Amsterdam, Netherlands\\
              \email{f.a.kummer@uva.nl}
              \and 
              Institute of Astronomy, KU Leuven, Celestijnenlaan 200D, 3001 Leuven, Belgium\\ 
             }

   \date{Received 1 August 2025 / Accepted 11 September 2025}

  \abstract
   {Massive stars are often born in triples, where gravitational dynamics and stellar interactions play a crucial role in shaping their evolution. One such pathway includes the merger of the inner binary, transforming the system to a binary with a distinct formation history. Therefore, the interpretation of observed binary properties and their inferred formation history may require the consideration of a potential triple origin.}
   {We aim to investigate the population of stellar mergers in massive hierarchical triples. Specifically, we assess how frequently mergers occur, and characterise the properties of the post-merger binaries and their subsequent evolution.}
   {We combine the triple population synthesis code {\tt TRES}, which self-consistently models stellar evolution, binary interaction, and gravitational dynamics, with the binary population synthesis code {\tt SeBa} to simulate $10^5$ dynamically stable, massive triples from the zero-age main sequence through merger and post-merger evolution. We explore the effects of a range of physical models for the initial stellar properties, mass transfer, and merger.}
   {We find that stellar mergers are a common outcome, occurring in 20–32\% of massive triples. Most mergers happen relatively early in the evolution of the system and involve two main-sequence (MS) stars, producing rejuvenated merger remnants that can appear significantly younger than their tertiary companions. Consequently, we predict that 2-10\% of all wide MS+MS binaries ($P>100$ days) have a measurable age discrepancy, and serve as a promising way to identify merged stars. The post-merger systems preferentially evolve into wide, eccentric binaries, with $\sim$80\% avoiding further interaction. However, a notable fraction (16–22\%) undergoes a second mass-transfer phase, which may result in the formation of high-mass X-ray binaries or mergers of compact objects that spiral in via gravitational-wave emission. Our results highlight the crucial role that stellar mergers in triples play in shaping the population of massive binary stars.}
   {}


   \maketitle
%

\section{Introduction}

Triple systems are common within the population of massive stars. For stars with masses above $8\,\rm{M}_{\odot}$, more than 50\% of young systems in the Galactic field are observed to be triples or higher-order multiples \citep{moe2017,offner2023,frost2025}. In most of these systems, stellar evolution and secular three-body dynamics drive the inner binary toward mass transfer \citep{stegmann2022a,kummer2023}. Observational studies further show that a large fraction of very short-period massive binaries reside in triple or higher-order multiple systems \citep{abdul-masih2025}. If during the mass transfer sufficient angular momentum is carried away, it can lead to orbital shrinkage and ultimately a stellar merger. Earlier binary population synthesis studies suggest that mergers are a common outcome of binary evolution \citep[e.g.,][]{demink2014,toonen2017,temmink2020}.

Identifying stellar mergers observationally remains challenging. Most predicted signatures can not unambiguously be attributed to mergers. For example, red novae are often interpreted as observational manifestations of stellar mergers, but they are also attributed to thermonuclear outbursts \citep{soker2003,tylenda2006,tylenda2011,kochanek2014,pejcha2014,smith2016}. Such transient events are rare due to the brief duration of the merger process.

Merger remnants may retain signatures of their merger history. For up to about $10^4$ years, the post-merger star is expected to be surrounded by a nitrogen-enriched, bipolar nebula as a result of mass loss \citep{langer2012}. While such nebulae may also arise from other forms of binary interaction, some Galactic nebulae have morphologies consistent with expectations from massive stellar mergers \citep{leitherer1987,smith2018}. Additional evidence comes from stellar surface properties: main-sequence (MS) merger remnants are expected to rotate slowly and generate large-scale magnetic fields \citep{schneider2016,schneider2019}. About 8\% of the massive stars has a detected magnetic field \citep{hubrig2008,wade2015,scholler2017}, although it remains debated whether such fields can also originate from primordial mechanisms. Massive merged stars may also show peculiar surface chemical abundances, such as CNO-cycle products \citep[e.g.,][]{glebbeek2013,patton2025}.

If a merger occurred in a triple system, the outer companion may provide indirect but valuable clues to the merger history. For instance, \citet{heintz2022,heintz2024} found that in about 20\% of the wide binary white dwarf (WD) systems, the components have an apparent age discrepancy, suggestive of prior stellar interaction. Given their wide orbits, it is unlikely that the stellar progenitors have engaged in mass transfer, favouring a merger scenario. \citet{shariat2025} performed a theoretical population study of low- and intermediate mass triples, and showed that the binary WD systems formed following a stellar merger are consistent with the observed properties of wide non-coeval WD binaries. 
Similarly, \citet{frost2024} recently reported an age discrepancy between the massive magnetic star in HD 148937 and its wide MS companion, consistent with a past merger event.

While the role of stellar mergers in low- and intermediate-mass triples has been theoretically explored in some detail \citep{shariat2025}, such a study of stellar mergers in massive triple systems is still lacking. \citet{stegmann2022b} investigated the formation of binary black hole (BBH) mergers from massive triples, but did not focus on the intermediate evolutionary stages and their observational characteristics.

In this study, we explore the population of stellar mergers in massive hierarchical triple systems in the Galaxy, including their post-merger evolution. We examine how the tertiary companion may aid in identifying these mergers observationally. To this end, we carry out simulations using a triple population synthesis approach. In Sect. \ref{sec:sec2}, we describe the methodology used to model triple evolution and mergers. In Sect. \ref{sec:sec3}, we present the properties of the mergers, various evolutionary pathways they follow, and zoom in on a few interesting subpopulations. In Sect. \ref{sec:sec4}, we provide merger rate predictions for the Galaxy and discuss limitations of the model. Finally, Sect. \ref{sec:sec5} summarises our findings.

\section{Methods}
\label{sec:sec2}
To investigate the population of stellar mergers within massive hierarchical triple systems, we simulated the evolution of initially high-mass ($m >8\,\rm{M}_{\odot}$), dynamically stable triples. First, we modelled the triple evolution from the zero-age main sequence (ZAMS) up to the onset of mass transfer in the inner binary using a triple evolution code. Once mass transfer commences, we modelled this interaction using a binary population synthesis code. If the mass transfer leads to a merger, we continue to follow the subsequent evolution of the resulting binary, now composed of the merger remnant and the original tertiary companion, within the same framework. In this section, we describe the methods applied in each phase of the simulation, followed by an overview of the initial conditions chosen for the synthetic triple population.   

\subsection{Triple evolution}
To model the evolution of massive hierarchical triple systems up to the onset of the first mass-transfer phase, we used the publicly available triple population synthesis code {\tt TRES}\footnote{\url{https://github.com/amusecode/TRES}} \citep{toonen2016}. This code self-consistently incorporates single-star evolution, binary interactions, and secular three-body dynamics. Stellar evolution is modelled using the stellar evolution code {\tt SeBa} \citep{portegies1996,toonen2012}, that includes analytic fits of \citet{hurley2000} to pre-calculated stellar models from \citet{pols1998}, allowing for fast calculation of stellar properties such as the mass, radius, and effective temperature, though not the internal structure. The strength of this code lies in its ability to model large stellar populations efficiently. We note that the analytical fits for single-star evolution become less accurate toward higher initial masses, especially at $m>50\,\rm{M}_{\odot}$, resulting in an overestimation of the maximum stellar radius \citep{shariat2025,sciarini2025}, and underestimation of the final compact object (CO) mass \citep{bavera2023} for the most massive systems.

A hierarchical triple is defined as a system with an inner binary composed of a primary ($m_1$) and a secondary star ($m_2$), orbited by a more distant tertiary star ($m_3$). The motion of the tertiary star around the centre of mass of the inner binary is referred to as the outer orbit. Such systems are dynamically stable on stellar evolutionary timescales when sufficiently hierarchical. The relevant elements that describe the shape of the inner and outer binary are the semi-major axes ($a_{\text{in}}$, $a_{\text{out}}$), eccentricities ($e_{\text{in}}$, $e_{\text{out}}$), arguments of pericentre ($g_{\text{in}}$, $g_{\text{out}}$), and the mutual inclination $i_{\text{mut}}$ between orbital planes. 

The orbital properties evolve through several processes, including tidal dissipation, gravitational-wave (GW) emission, three-body dynamical interactions, and stellar winds. These effects are orbit-averaged and solved using a set of first-order ordinary differential equations within {\tt TRES}. Tidal evolution follows the formalism of \citet{hurley2002}, where the equilibrium and dynamical tides are based on results of \citet{hut1981} and \citet{zahn1997}, respectively. Equilibrium tides are applied to stars with convective envelopes, while dynamical tides are applied to stars with radiative envelopes. GW-driven orbital evolution follows \citet{peters1964}. Precession from tidal, rotational, relativistic, and dynamical processes is included. The secular three-body interactions are treated through the quadrupole and octupole-order expansions of the Hamiltonian \citep{vonzeipel1910,lidov1962,kozai1962,naoz2016}, capturing modulations such as von Zeipel-Lidov-Kozai (ZLK) oscillations. For the details on the implementation and the effects of these processes, we refer to \citet{toonen2016}.

For wind mass loss of stars on the MS, we follow \citet{vink2001} for effective temperatures between $\text{8-50\, kK}$, and \citet{nieuwenhuijzen1990} outside this range. These mass-loss rates are scaled down by a factor of three, in accordance with \citet{bjorklund2021}. For Hertzsprung Gap (HG) and core helium burning (CHeB) stars, we use the same prescription, but compare it with the \citet{reimers1975} rate, and adopt the maximum of the two. For asymptotic giant branch (AGB) stars, the maximum of \citet{nieuwenhuijzen1990}, \citet{reimers1975} and \citet{vassiliadis1993} is used. For stripped helium stars, we follow \citet{hamann1995} and \citet{hamann1998}, including a metallicity scaling of $Z^{0.86}$ \citep{vink2001}. For stars evolving beyond the Humphreys-Davidson limit, an additional mass-loss rate of $1.5\times10^{-4}\, \text{M}_{\odot}\text{yr}^{-1}$ is included \citep{belczynski2010}.

When a massive star forms a CO at the end of its evolution, eventual mass loss during the supernova (SN) may lead to additional changes in the orbital separation and eccentricity. Assuming asymmetric mass ejection, we follow the distribution of kick velocities presented in \citet{verbunt2017}, which is scaled down for black holes (BHs) based on their mass. Natal kick velocities for BHs are uncertain, and observations of BHs in multiple systems suggest their formation may be accompanied by little or no kick \citep[e.g.,][]{shenar2022,vigna-gomez2024,burdge2024,shariat_xray2025,vanson2025}. The mass of the CO is determined following the delayed SN model from \citet{fryer2012}, which maps the final carbon-oxygen core mass to the remnant mass.

Each triple was evolved with {\tt TRES} from the ZAMS until the primary star fills its Roche lobe, ensuring that neither component of the inner binary has evolved into a CO. Simulations were terminated and the system discarded if any of the following conditions are met: the tertiary star fills its Roche lobe, the inner or outer orbit becomes unbound (e.g., due to a supernova kick), the system becomes dynamically unstable based on the stability criterion of \citet{mardling1999,mardling2001}, the runtime exceeds three hours (to ensure computational feasibility applicable for systems with very short dynamical timescales relative to stellar evolution, although this limitation likely excludes some potentially interesting cases with strong three-body dynamics), or no interaction occurs within a Hubble time, including mass transfer in the inner binary. 

\subsection{Mass transfer}

Mass transfer is initiated when the primary star fills its Roche lobe. This phase is modelled separately with {\tt SeBa}. At the onset of Roche-lobe overflow (RLOF), the orbit is not necessarily circular. This can be due to prior dynamical interactions that increase the orbital eccentricity, or because tidal forces have not yet circularised the system. Therefore, we define the Roche radius at pericentre:
\begin{equation}
\label{eq:roche_lobe}
    R_{\text{L,1}} = \frac{0.49q^{2/3}}{0.6q^{2/3}+\text{ln}(1+q^{1/3})}a(1-e),
\end{equation}
where $q=m_1/m_2$ is the mass ratio, $a$ is the semi-major axis, and $e$ is the orbital eccentricity. Since a comprehensive analytical prescription for mass transfer that is valid for all eccentricities, mass ratios, and stellar spins is still lacking, despite significant progress in recent studies \citep{sepinsky2007,spinsky2009,hamers2019}, we assumed the binary rapidly circularises at the onset of the mass transfer.

The stability of mass transfer depends on the specific conditions of the binary at the onset of RLOF. To assess this, follow the approach of SeBa. We compare the donor star's response to mass loss, both adiabatic and thermal, to the evolution of the Roche lobe, which changes in size as the semi-major axis and mass ratio vary during the mass transfer. This response is characterised as the logarithmic derivative of the stellar radius or Roche-lobe radius with respect to the mass:
\begin{equation}
\zeta\,=\,\bigg(\frac{\text{d\,ln}R}{\text{d\,ln}M}\bigg).
\label{eq_zeta}
\end{equation}
If the Roche-lobe radius shrinks faster than the donor's radius when restoring hydrostatic equilibrium (i.e., $\rm{\zeta_{ad} < \zeta_L}$), mass transfer becomes dynamically unstable and leads to a common envelope (CE) phase. Otherwise, the mass transfer is stable and proceeds on either the nuclear timescale $(\rm{\zeta_L \leq min(\zeta_{ad}, \zeta_{th})})$ or the thermal timescale $\rm{(\zeta_{th} \leq \zeta_{L} \leq \zeta_{ad})}$ of the donor star, depending on whether the donor star can re-establish thermal equilibrium or not.  

For dynamically unstable systems, we modelled the subsequent CE evolution using the standard $\alpha\lambda$ prescription \citep{webbink1984,dekool1990}, commonly applied to massive stars. Here, $\alpha$ represents the efficiency with which orbital energy is used to eject the envelope, while $\lambda$ represents the binding energy of the envelope and depends on its internal structure. For systems undergoing CE evolution, we assumed a fixed value of $\alpha\lambda = 2$ \citep{Nelemans2000} in our fiducial model. If the envelope cannot be fully ejected before the core of the donor and the accretor come into contact, the binary is assumed to merge.

During a phase of stable mass transfer, angular momentum is redistributed within the binary, and leads to a change in the orbital separation. If mass transfer is conservative, the orbit initially shrinks while the donor is more massive than the accretor, and then widens once the mass ratio reverses. In non-conservative mass transfer, the orbital evolution depends on the specific angular momentum carried away by the lost material. How much mass is lost depends on the amount of mass the accretor is able to accrete. We determined the maximum accretion rate using the prescription from \citet{pols1994}, and assumed that any escaping material carries away angular momentum equal to 2.5 times the specific orbital angular momentum \citep{portegies1996}. If the orbit shrinks to the point where the accretor fills its own Roche lobe, the system enters a contact phase. For simplicity, we assume that stars in contact immediately merge, although this outcome is not guaranteed \citep[see e.g.,][]{henneco2024}. 

We accounted for the effect of mass loss from the inner binary on the outer orbit. Following the approach of \citet{shariat2023}, if the time interval on which mass loss occurs is longer than the orbital period of the outer binary, we assumed it to be isotropic and adiabatic. By approximation, the lost mass carries a specific angular momentum equal to the specific angular momentum of the inner binary. Under these conditions, the outer orbit evolves as:
\begin{equation}
    a_{\text{out,f}} = a_{\text{out,i}}\Bigg(\frac{m_{\text{binary,i}}}{m_{\text{binary,f}}}\Bigg)^{2}\frac{m_{\text{triple,i}}}{m_{\text{triple,f}}},
\end{equation}
where the subscripts "binary", and "triple" refer to the combined masses of the inner binary and the full triple system, respectively, and the subscripts "i" and "f" denote the initial and final values. If the mass loss occurs on a timescale shorter than the outer period, e.g., during a CE phase, we treat it as instantaneous. In this case, a velocity kick is applied to the outer orbit \citep[see][]{hills1983}, which alters both the semi-major axis and the eccentricity \citep[e.g.,][]{michaely2019,igoshev2020}.

\subsection{Stellar mergers and post-merger evolution}

The outcome of a stellar merger in our models depends on the evolutionary stages and (core) masses of the stars involved. For mergers between two MS stars, the mass of the accretor is added fully conservatively to the envelope of the donor, resulting in a more massive MS star \citep[e.g.,][]{glebbeek2013}. Since higher mass stars have shorter MS lifetimes, the merger remnant appears younger than its true age. Additionally, the core-to-envelope mass ratio decreases as material is added to the envelope, contributing further to the rejuvenated appearance. To capture both effects, the rejuvenation due to envelope accretion determines the relative age of the merger remnant, following the prescription of \citet{hurley2002}:
\begin{equation}
\label{eq:rejuvenation}
    t_{\rm{rel,f}} = t_{\rm{rel,i}}\frac{\tau_{\rm{ms,f}}}{\tau_{\rm{ms,i}}}\frac{m_{\rm{i}}}
    {m_{\rm{f}}},
\end{equation}
where $\tau_{\rm{ms}}$ is the MS lifetime, and the mass ratio accounts for the change in stellar mass. We did not include mixing of hydrogen into the core during the merger, which would lead to additional rejuvenation \citep[e.g.,][]{lombardi2002,glebbeek2008,glebbeek2013}. We also assumed no mass loss during the merger in our fiducial model, which is broadly consistent with findings of \citet{glebbeek2013}, who found that no significant mass loss is expected for MS mergers. If the donor is a post-MS star, the accretor is added fully conservative to the envelope of the donor. In cases where both stars are post-MS stars, we assumed the cores merge conservatively, but half of the envelope mass is lost during the merger \citep{portegies1996}. See Sect. \ref{sec:model_variations} for model variations on the merger physics.

We do not account for other merger-induced processes that may affect subsequent stellar evolution, such as rotationally enhanced mixing and mass loss, magnetic braking due to internal field generation \citep{schneider2016,schneider2019}, bloating \citep{suzuki2007,stegmann2022b} or changes to the core structure.

After the merger, the evolution of the resulting post-merger binary (i.e., the merger remnant and the remaining third star) is followed with {\tt SeBa} using the same physics as applied during the earlier triple evolution and interaction phases. Each system is evolved until a Hubble time.

\subsection{Initial population}

In this section, we give an overview of the initial parameter distributions used to create our simulated population. We evolved a total of $10^5$ initially stable, hierarchical triple systems. The inner and outer semi-major axes, $a_{\text{in}}$ and $a_{\text{out}}$, were drawn from a distribution uniform in logarithmic space ranging from 5 to $5\times10^6 \, \rm{R_{\odot}}$, in broad agreement with \citet{kobulnicky2007} and \citet{sana2012}. If the drawn value of $a_{\text{in}}$ exceeded $a_{\text{out}}$, the two values were swapped to ensure $a_{\text{in}} < a_{\text{out}}$. The eccentricities $e_{\text{in}}$ and $e_{\text{out}}$ were sampled from a thermal distribution between 0 and 1 \citep[e.g.,][]{moe2017,hwang2023,shariat_gaia2025}. If either or both stars in the inner binary initially filled their Roche lobe, lower values of $e_{\text{in}}$ were redrawn until the system became detached. If the system was not detached within ten attempts, the triple was discarded. The inner and outer mass ratios, defined as $q_{\text{in}} = m_2/m_1$ and $q_{\text{out}} = m_3/(m_1+m_2)$, were sampled from a uniform distribution between 0 and 1  \citep{sana2012,kobulnicky2014}. Note that the definition of $q_{\rm{in}}$ is the reciprocal of the definition given in Eq. \ref{eq:roche_lobe}. The arguments of pericentre, $g_{\text{in}}$ and $g_{\text{out}}$, were drawn randomly between 0 and $2\pi$, and the mutual inclination $i_{\text{mut}}$ was sampled randomly from a distribution uniform in $\cos{i}$ between 0 and $\pi$ \citep[e.g.,][]{shariat_gaia2025}. We note that systems with compact outer binaries seem to favour more aligned orbits \citep[e.g.,][]{borkovits2015,tokovinin2017,bashi2024}, which would diminish the importance of three-body dynamics for such systems. 

\begin{figure}
    \includegraphics[width=\hsize]{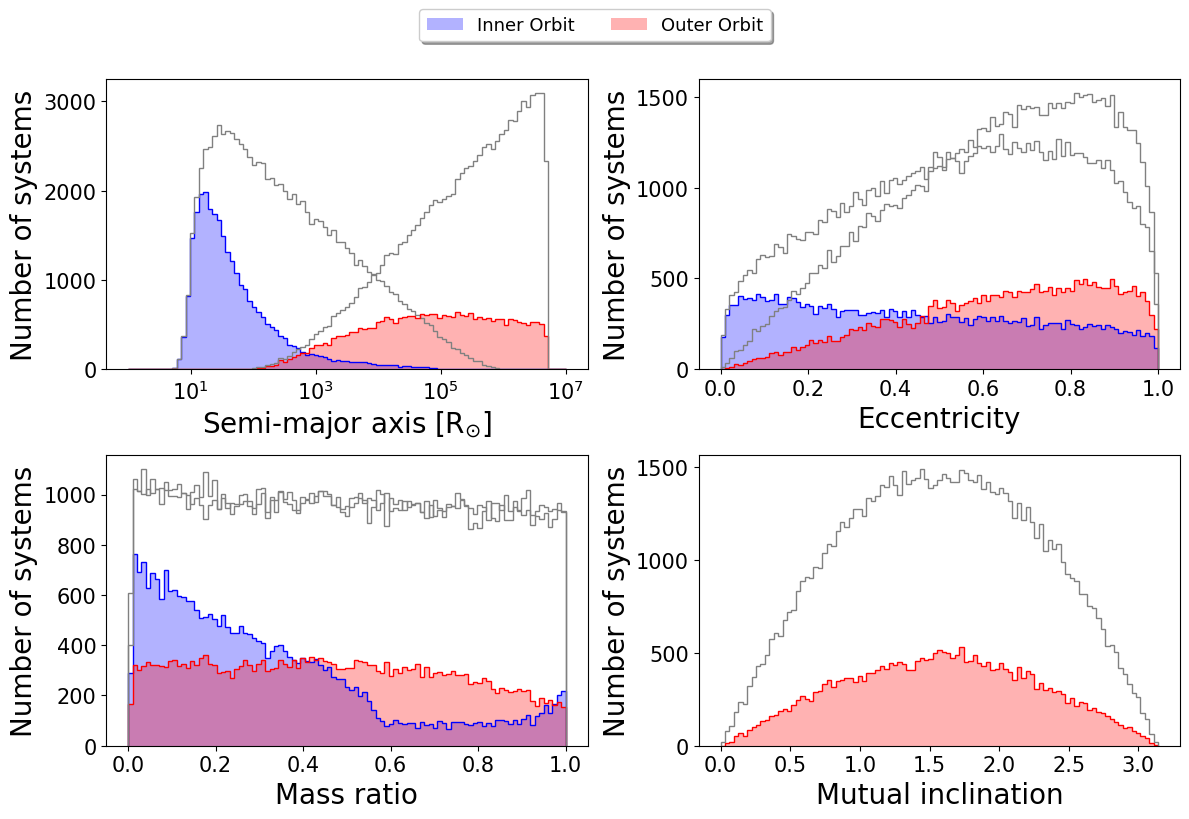}
    \caption{Distributions of the initial (ZAMS) orbital properties for systems that result in a stellar merger in the fiducial model. On the y-axis, we show the number of simulated systems. Properties of the inner orbits are shown in blue, outer orbits in red, and the complete simulated population is shown in gray for comparison. Panels show (top left) semi-major axis, (top right) eccentricity, (bottom left) mass ratio ($q = m_2/m_1$ for inner orbits, $q = m_3/(m_1+m_2)$ for outer), and (bottom right) mutual inclination between the inner and outer orbits.}
    \label{fig:figure1}
\end{figure}

Systems found to be dynamically unstable at initialisation were discarded and redrawn. Also, we excluded systems that initiated mass transfer within $10^5\,\rm{yr}$ from our analysis, which comprise about 5\% of the population. These systems often undergo strong dynamical interactions and would likely have interacted already during the pre-MS phase. Consequently, the final distribution of the ZAMS triples differs from the initial sampling distributions. The modified ZAMS distribution is marked by the gray histogram in Fig. \ref{fig:figure1}. 

\subsection{Model variations}
\label{sec:model_variations}

A major challenge in population synthesis studies of multiple stellar systems is the large number of assumptions required to model their evolution. These assumptions reflect uncertainties in both the initial conditions and the complex physical processes involved in their evolution (e.g., mass transfer physics, wind mass loss, SN kick and engine). Specifically, simulations that extend to the late stages of stellar evolution can produce widely varying outcomes depending on the adopted assumptions \citep[e.g.,][]{broekgaarden2021,broekgaarden2022}. There are many assumptions to probe, but we focus on a few to which we anticipate that the outcomes are particularly sensitive to, namely, uncertainties related to the initial conditions, mass transfer and the merger. To explore the sensitivity of our results to these uncertainties, we ran several model variations, each modifying a single aspect of the fiducial setup. 

\paragraph{Initial triple properties:}
We varied the outer mass ratio distribution, which is poorly constrained observationally. Previous work \citep[e.g.,][]{kummer2023} has shown that the mass of the tertiary has a notable influence on the overall evolution of massive triple systems, and that the distribution of other properties are of less importance. In this model variation, we simulated an additional $10^5$ massive triple systems, still sampling the outer mass ratio from a uniform distribution, but redefining it as $q_{\rm{out}} = m_3/m_1$, which results in initially lower-mass tertiary stars compared to the default definition $q_{\rm{out}} = m_3/(m_1+m_2)$. This tendency toward lower-mass tertiaries is consistent with observations of early B- and O-type primaries \citep{moe2017}.

\paragraph{Mass transfer:}
The physics of mass transfer remains one of the most uncertain aspects of binary and multiple star evolution. We tested a model assuming fully conservative stable mass transfer, leading to more mass being retained within the system. We also varied our treatment of the CE phase. Instead of the default assumption of $\alpha\lambda = 2$, we ran a model with $\alpha\lambda = 0.25$, motivated by observations of post-CE binaries \citep[e.g.,][]{zorotovic2010,toonen2013,camacho2014}, which suggests a less efficient envelope ejection. Additionally, we tested a model in which the mass loss from a CE event during pre-merger mass transfer occurs on a longer timescale of $10^4 \, \text{yr}$. Observations of post-CE binaries and triples suggest that, in some cases, mass ejection may be slower than typically assumed \citep{michaely2019,igoshev2020,knigge2022,shariat_cv2025}. We note that these studies focus on low- to intermediate-mass stars.

\paragraph{Merger:}
We explored the effect of including additional mass loss during the merger process itself. For MS+MS mergers, we adopted the relation from \citet{glebbeek2008}:
\begin{equation}
    \phi = 0.3\frac{q_{\rm{in}}}{(1+q_{\rm{in}})^2},
\end{equation}
where $\phi$ represents the fraction of the total mass lost during the merger. For mergers involving a post-MS donor and a MS accretor, we treated them analogously to post-MS + post-MS mergers, assuming that half of the envelope mass is lost during the event. For the variations of the mass transfer and merger physics, we did not re-simulate the triple systems with the triple evolution code, but only continued their evolution from the onset of mass transfer in the inner binary.

\section{Results}
\label{sec:sec3}
\subsection{Mergers: statistics and properties}

Across all model variations, we find that stellar mergers are a common outcome in the evolution of massive hierarchical triple systems. Depending on the adopted initial conditions and assumptions about mass transfer and merger physics, the merger fraction ranges from 20.4\% to 31.8\% of all simulated systems. These fractions are broadly consistent with those reported in \citet{stegmann2022a} and \citet{preece2024}, who find that 15-33\% and 38\% of the systems undergo a stellar merger, respectively. This high incidence of mergers is mainly a result of the short initial orbital periods of the inner binaries. These compact configurations are necessary to ensure dynamical stability within the triple system, and as a consequence, there is a high rate of mass transfer interactions \citep{stegmann2022a,kummer2023}, facilitating the occurrence of mergers.

\begin{figure}
    \includegraphics[width=\hsize]{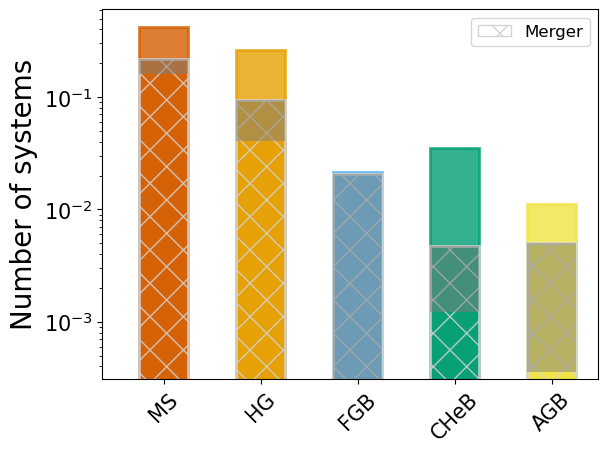}
    \caption{Distribution of the evolutionary phases of donor stars at the onset of mass transfer for systems in the fiducial model. On the y-axis, we show the number of simulated systems. Each coloured bar corresponds to a different evolutionary phase: Main Sequence (MS), Hertzsprung Gap (HG), First Giant Branch (FGB), Core Helium Burning (CHeB), and Asymptotic Giant Branch (AGB). The hatched gray overlay indicates the fraction of systems in each category that result in a stellar merger. Error bars (shaded regions) represent the range of values for the stellar mergers across all model variations.
    }
    \label{fig:figure2}
\end{figure}

The majority of mergers (63-79\%) occur between two MS stars, followed by mergers of a HG donor and a MS star (20-30\%), as illustrated in Fig. \ref{fig:figure2}. The contribution of other donor types is less than 10\% combined. We did not consider mergers containing a neutron star (NS) or BH companion, as such systems would be dominated by mergers that occur during the second mass-transfer phase, which is not modelled in this study. In contrast to low-mass stars, massive stars undergo significant expansion during the MS phase, increasing the likelihood of RLOF before the donor evolves beyond the MS. As a result, short-period inner binaries often initiate mass transfer during or shortly after the MS phase of the donor. Among the explored models, the largest variation in merger outcomes is introduced by the efficiency of the CE phase. Less efficient expulsion of the envelope leads to a substantial increase in the number of post-MS+MS mergers, as the mass transfer in such systems is typically unstable. In contrast, the number of MS+MS mergers varies more moderately, from 63\% to 79\%.

\begin{figure}
    \includegraphics[width=\hsize]{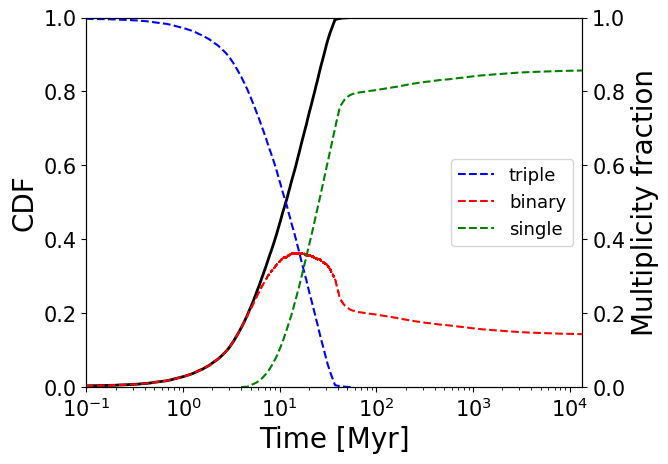}
    \caption{Fraction of merging triple systems that have experienced a merger of the inner binary as function of time for the fiducial model (black solid line). The dashed lines represent the fraction of systems --- from the population that at some point in time experience a stellar merger of the inner binary --- that contain a bound triple (blue), a binary (red), or only single stars (green) at a given time.}
    \label{fig:figure3}
\end{figure}

The timing of mergers is further detailed in Fig. \ref{fig:figure3}, considering only the systems that undergo a merger. Mergers occur predominantly early, with the rate rising sharply after 2–3 Myr and nearly 50\% of the systems having merged by 10 Myr. All mergers occur within 55 Myr, which corresponds roughly to the lifetime of an $8 \,\rm{M}_{\odot}$ star --- the lowest-mass primary star considered in our population. If lower-mass stars were included in the simulations, more mergers would occur later on. As time progresses, triple systems gradually disappear, while the number of binaries increases. However, post-merger interactions can further disrupt these binaries, leading to single stars. These interactions are addressed in the later sections of this paper. The binary fraction remains high for an extended period of time. For the fiducial model, the binary fraction is above 0.3 after 7-40 Myr, suggesting that binaries containing merger remnants are potentially observable over extended periods. Depending on the model variations, the binary fraction reaches maximum values of 0.3-0.43. A long-lived, slowly decreasing tail in the binary distribution persists out to and beyond a Hubble time, primarily tracing GW-driven double CO (DCO) mergers. For an overview on the impact of other evolutionary channels and higher-order stellar configurations on stellar multiplicity, we refer the reader to \citet{preece2024}.

In Fig. \ref{fig:figure1}, we explore the initial properties of merging systems. In blue (red) the ZAMS parameters of the inner (outer) orbit for triples with merging inner binaries are shown. These binaries preferentially have short orbital periods and low eccentricities, consistent with tighter binaries being more prone to mass transfer. The bias toward low eccentricities for short-period inner binaries arises from redrawing the initial eccentricity whenever the inner binary would otherwise be Roche-lobe filling at birth. We also find a tendency toward unequal mass ratios, which both increase the likelihood of unstable mass transfer (driving rapid inspiral) and enhance orbital shrinkage during stable mass transfer episodes following from the transfer and loss of angular momentum within the binary. Even though equal mass ratios are disfavoured, a small peak is visible near $q_{\rm{in}}=1$, corresponding to cases where both stars evolve on similar timescales. There is a small dependence on the mutual inclination, with a modest peak around $i_{\rm{mut}}=90^\circ$. In the fiducial model, about 7\% of the inner binaries that experience a stellar merger experience an eccentricity increase of at least 0.1 due to three-body dynamics, a trend that is reflected in the inclination distribution. This suggests that while three-body dynamics can enhance the occurrence of mergers, it is not the dominant driver.

\begin{figure}[htbp]
  \centering
  \begin{subfigure}{\linewidth}
    \includegraphics[width=\linewidth]{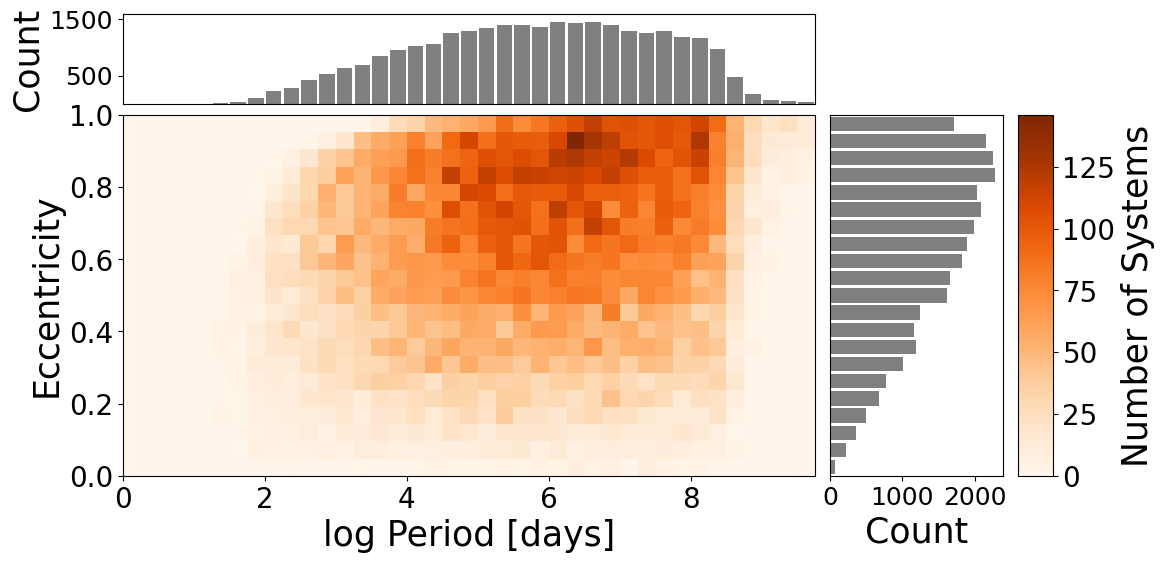}
  \end{subfigure}

  \begin{subfigure}{\linewidth}
    \includegraphics[width=\linewidth]{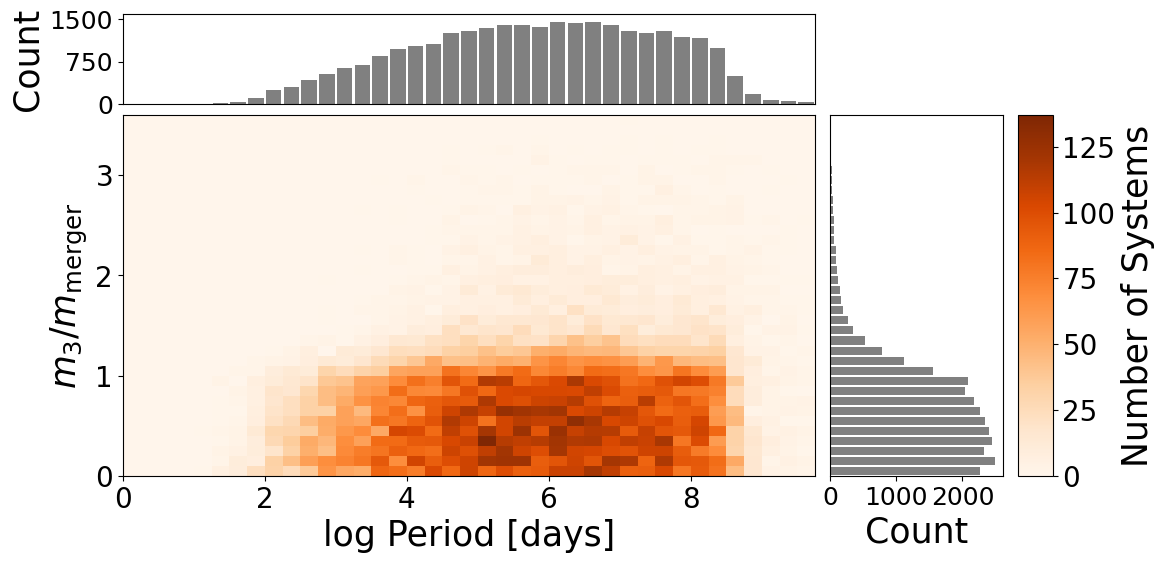}
  \end{subfigure}

  \begin{subfigure}{\linewidth}
    \includegraphics[width=\linewidth]{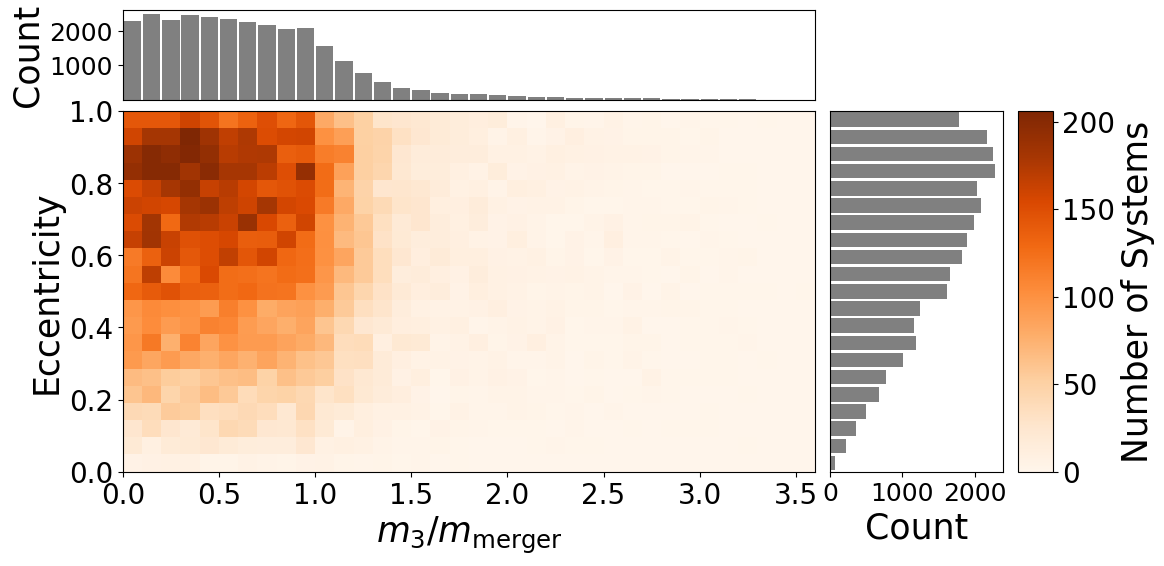}
  \end{subfigure}

  \caption{Shown are 2D histograms of the post-merger orbital parameters: log orbital period (in days) vs. eccentricity (top), log orbital period vs. tertiary-to-merger-remnant mass ratio $q_{\rm{pm}}$ (middle), and mass ratio vs. eccentricity (bottom). Colour indicates the number of simulated systems in each bin, and histograms on top and side axes show the marginal distributions.}
  \label{fig:figure4}
\end{figure}

After the merger, the orbital properties of the remaining binary system (i.e., the merger remnant and the tertiary) closely reflect those of the initial outer binary. Therefore, uncertainties in the initial assumptions for the outer binary will have a large impact on the post-merger orbits. Nevertheless, periods below 100 days are rarely expected, as a consequence of requiring dynamical stability at formation. In Fig. \ref{fig:figure4}, we show the orbits for the fiducial model just after the merger. The orbits are skewed toward ultra-wide periods, peaking from $10^5$ up to $10^8$ days, and the eccentricities approximately follow a thermal distribution. The post-merger mass ratio distribution, $q_{\rm{pm}} = m_3/m_{\rm{merger}}$, is roughly uniform between 0 and 1, with a tail extending to higher mass ratios. For the fiducial model, 20\% of the systems have a tertiary that is more massive than the merger remnant. This varies between 0-27\% based on the model assumptions. Even though we assume that the tertiary is initially never more massive than the inner binary, mass loss prior to and during the merger process can result in mass ratio reversal. We identify no significant correlations among period, eccentricity, and mass ratio in these systems.

\subsection{Post-merger evolution channels}
\label{sect:post_merger}
\begin{figure}
    \includegraphics[width=\hsize]{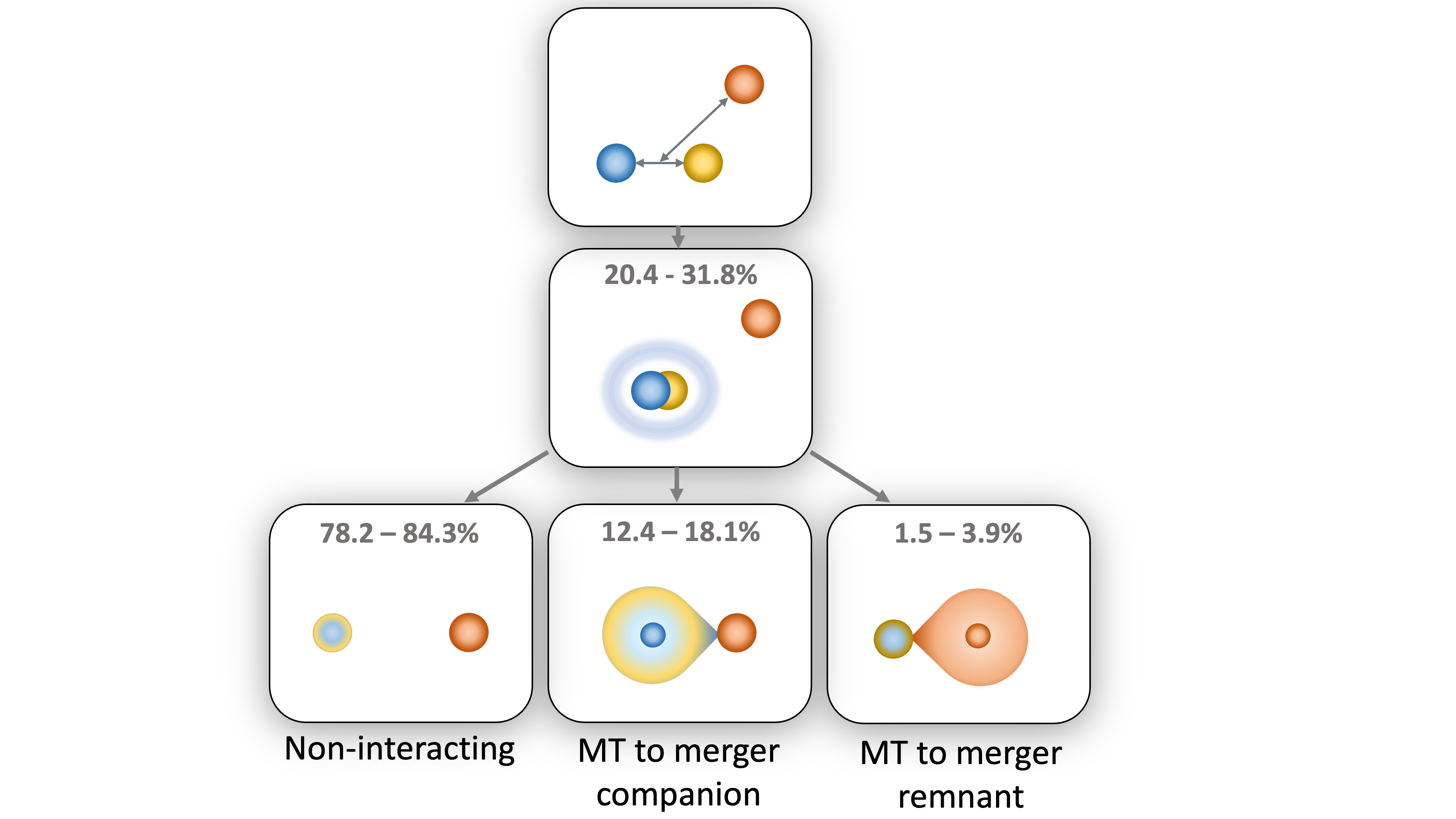}
    \caption{Illustration of the evolutionary channels of massive triples that experience a stellar merger with their respective incidences. The system initially forms as a stable hierarchical triple and undergoes a stellar merger within the inner binary as a result of mass transfer. The subsequent evolution of the post-merger binary proceeds along one of three possible channels, depending on the system's properties at the time of merger. Mass transfer is abbreviated as MT.}
    \label{fig:figure5}
\end{figure}

In this section, we outline the subsequent evolution of the binary system following the stellar merger, up to the end of the first mass-transfer phase. We classify this evolution into three distinct channels:
\begin{itemize}
    \item Non-interacting: systems in which no mass transfer occurs after the merger event.
    \item  Mass transfer to the merger companion: the merger remnant eventually fills its Roche lobe and transfers mass to the original tertiary star.
    \item Mass transfer to the merger remnant: the tertiary star fills its Roche lobe and initiates mass transfer onto the merger remnant.
\end{itemize}
A schematic illustration of these evolutionary channels is presented in Fig. \ref{fig:figure5}, and a summary of their occurrence rates can be found in Table \ref{table:merger_fraction}.

\subsubsection{Non-interacting systems}

The vast majority of post-merger binaries do not undergo further interaction (78.2–84.3\%). These systems have orbital periods of several thousand days or more, which prevents either component from ever filling its Roche lobe. Apart from the dearth of short periods, their properties closely resemble those of the overall post-merger population shown in Fig. \ref{fig:figure4}. Specifically, they tend to have eccentric orbits and mass ratios that are approximately uniformly distributed over the range $0 < q_{\rm{pm}} < 1$, with a tail extending toward more massive tertiary companions.

\subsubsection{Mass transfer to the merger companion}

In 12.4–18.1\% of post-merger binaries, a phase of mass transfer is initiated by the merger remnant. Their progenitors are predominantly MS+MS mergers, as the outer binary must have been sufficiently compact to allow for post-merger interaction. Consequently, the original inner binary must have been extremely tight to ensure dynamical stability before the merger. The resulting post-merger binaries have periods that peak sharply between one to several thousand days, and have eccentricities clustered around 0.8–0.9 directly after the merger. More extreme eccentricities at birth would often result in dynamical instability. The merger remnant is typically much more massive than the original tertiary; there is a sharp drop above $q_{\rm{pm}}=0.5$. Because MS+MS merger remnants are rejuvenated, in order for the merger remnant to leave the MS before the tertiary companion and be the first to fill its Roche lobe, in many cases it needs to be considerably more massive than the tertiary star.

\begin{figure}[htbp]
  \centering
  \begin{subfigure}{0.75\columnwidth}
    \includegraphics[width=\linewidth]{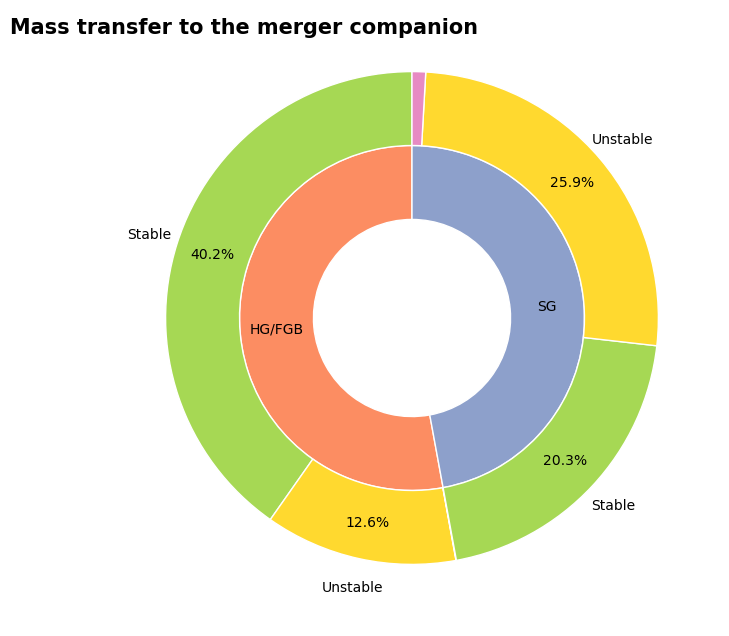}
  \end{subfigure}

  \begin{subfigure}{0.75\columnwidth}
    \includegraphics[width=\linewidth]{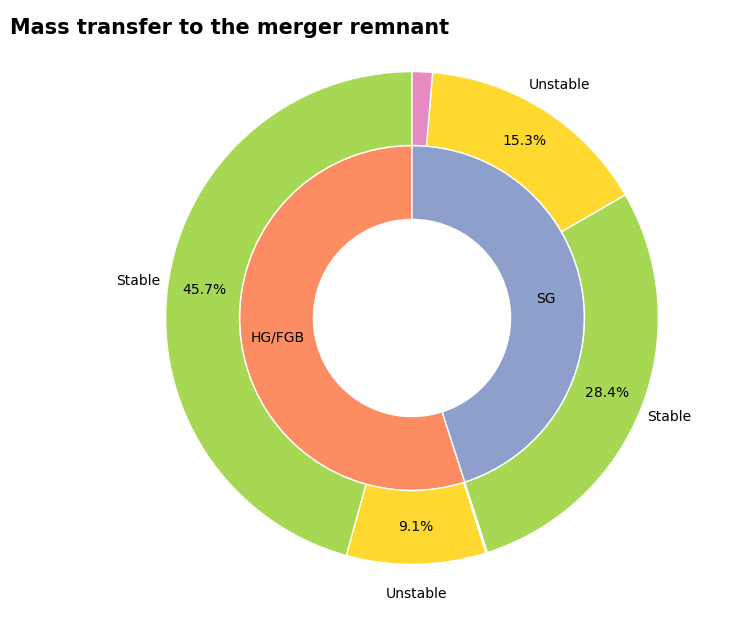}
  \end{subfigure}

  \caption{Donut chart showing the evolutionary state of the donor star (inner ring) and the stability of mass transfer (outer ring) at the onset of the second mass-transfer phase in the fiducial model. The top panel shows systems where the merger remnant becomes the donor, while the bottom panel shows systems where the original tertiary star initiates mass transfer. The inner ring distinguishes between systems that initiate mass transfer on the HG or FGB, and as a SG. The outer ring indicates whether the mass transfer is stable (green), unstable (yellow), or when the accretor is also a post-MS star (purple). The latter outcome is rare, comprising fewer than 2\% of systems.}
  \label{fig:figure6}
\end{figure}

The mass-transfer phase always happens when the donor star is on the post-MS, due to the lack of short period binaries. These are quite evenly split between systems with HG/FGB donors (52–58\%) and those with supergiant (SG) donors (42–48\%), as shown in Fig. \ref{fig:figure6}. As mentioned earlier, the stellar evolution tracks adopted in this work are known to overestimate the stellar radius, especially for initial masses above $50 \, \mathrm{M}_{\odot}$, which may lead to an overprediction of mass-transferring systems in that mass range. However, the impact of this effect is only minor due to the steeply declining initial mass function, which limits the contribution of the most massive stars to the overall population.

Mass transfer involving HG/FGB donors predominantly proceeds in a stable manner (73–80\%). In contrast, systems with more evolved donors --- those with deep convective envelopes --- more frequently undergo CE evolution (51–57\%). A small fraction ($<1\%$) of systems experience unstable mass transfer with a post-MS accretor, leading to a CE phase where both stars lose their envelope.

The properties of the post-mass-transfer orbits depend sensitively on the assumptions regarding the mass-transfer physics. In the case of fully conservative mass transfer, orbital periods are typically between 100 and 1.000 days, with tertiary masses peaking at $15 \, \mathrm{M}_{\odot}$ and corresponding mass ratios ($q_{\rm{pm}}$) around 3.5. Alternative models, which allow for non-conservative mass transfer, predict shorter post-mass-transfer periods due to loss of orbital angular momentum, ranging from 10 to 100 days, and tertiary masses peaking around $5 \, \mathrm{M}_{\odot}$, with mass ratios close to or just below unity.

Systems that come into contact undergo a second stellar merger, this time between the tertiary and the merger remnant, ultimately leaving behind a single star. We find that this outcome occurs in 9–31\% of systems with the merger remnant as donor, corresponding to 2–4\% of all triple mergers. The sequential mergers most commonly involve a hydrogen-shell burning or more evolved, (nearly) stripped donor and a MS accretor. Such sequential mergers may present intriguing targets for future work. 

\subsubsection{Mass transfer to the merger remnant}
\label{sec:mt_to_merger}
The least common evolutionary pathway (1.5–3.9\% of systems) involves cases where the original tertiary star fills its Roche lobe and initiates mass transfer toward the merger remnant. While the post-merger orbital periods and eccentricities in these systems resemble those where the merger remnant is the donor, their mass ratios are centred around unity. On the one hand, this trend arises because the tertiary has to evolve faster than the merger remnant and can thus not be much lower in mass originally (although due to rejuvenation of the merger remnant, the mass can be somewhat lower during the merger). On the other hand, the tertiary star should not fill its Roche lobe before the merger remnant does, and can thus not be of much higher mass. An additional limiting factor is that in our simulations, the tertiary star is initially never more massive than the inner binary. 

Mass transfer from the tertiary companion to the merger remnant is more likely to be stable than in cases where the merger remnant initiates transfer, despite the donors being in similar evolutionary phases (see Fig. \ref{fig:figure6}). This is mainly due to the differences in the mass ratio at the moment mass transfer is initiated. As a result, the orbital architectures after mass transfer differ significantly. Firstly, mass transfer from the tertiary leads to wider orbits. In the case of conservative mass transfer, orbital periods typically range between $10^3$–$10^4$ days, and the other model variations predict a peak around 1000 days. Secondly, the masses of the accretor star are larger, peaking slightly above $20 \, \rm{M}_\odot$. The resulting mass ratios are just below 0.1. Due to their wide orbits and tendency toward stable mass transfer, these systems rarely experience a second stellar merger, making the contribution of sequential mergers negligible. 

Interestingly, in approximately 10\% of the systems where the tertiary initiates mass transfer to the merger remnant, the merger remnant has already evolved into a CO. These cases originate from mergers involving post-MS donors that lose their envelopes during the initial mass transfer and merger event. As a result, the stripped merger remnant remains relatively compact for the remainder of its evolution, avoiding RLOF. However, due to the current lack of detailed theoretical models for post-MS mergers and associated envelope loss, we caution against overinterpreting these results.

\subsection{Focus on interesting systems}

Here we describe types of systems that are important in the context of mergers in triple evolution.

\subsubsection{Non-coeval binaries}

Accretion of material onto a star's envelope is expected to rejuvenate it, making the star appear younger compared to its true age. In multiple-star systems, rejuvenation can lead to observable age discrepancies between the rejuvenated star and its companion(s). Several processes can lead to this effect, such as mass transfer \citep{mccrea1964,chen2008,ivanova2015}, stellar mergers induced by binary interaction \citep{perets2009,naoz2014,shariat2025}, and direct stellar collisions \citep{hills1976,sills1997,sills1999,sills2005}. These processes are for instance used to explain the formation of blue straggler stars (BSSs) observed in coeval stellar clusters. 

Recent observational evidence adds support to the hypothesis that stellar mergers within triple systems contribute to the formation of non-coeval massive binary stars. \citet{frost2024} reported on a massive MS+MS binary in which one component appears at least 1.4 Myr younger than its companion. Given the binary’s wide orbit ($\geq$ 18 years), mass transfer is unlikely to have occurred, and the observed configuration is best explained by a merger of two MS stars within a former inner binary of a hierarchical triple. Moreover, the rejuvenated star shows clear signs of a strong magnetic field, while its companion does not. Theoretical models predict that mergers between MS stars can produce rejuvenated MS stars with large-scale magnetic fields \citep{schneider2016,schneider2019}.

\begin{figure}
\includegraphics[width=\hsize]{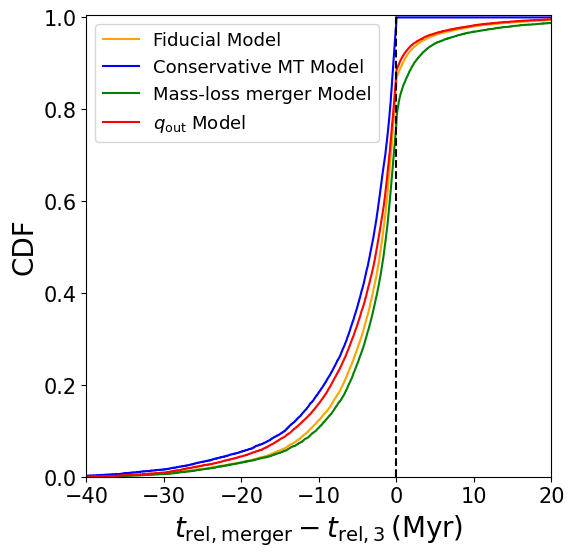}
    \caption{Difference in relative (apparent) age between the merger remnant and the original tertiary star directly after the merger of the inner binary, for four selected models. Negative values indicate systems in which the merger remnant appears younger (i.e., rejuvenated) than the original tertiary companion. Positive values correspond to systems where the merger remnant appears older. The black dashed line marks the transition between rejuvenated and older-appearing merger remnants. Only models that deviate significantly from the fiducial model are shown; those that are nearly identical are omitted for clarity.}
    \label{fig:figure7}
\end{figure}

In our simulated population, we investigate the degree of rejuvenation in MS+MS mergers with a MS tertiary. These systems dominate the overall merger population. In Fig. \ref{fig:figure7}, we show the distribution of apparent age difference between the merger remnant and the tertiary companion. The star's apparent age is determined by its position along the single-star evolution track corresponding to its current mass. Mass accretion can shift the star to change onto a different track, giving it a rejuvenated appearance, as described by Eq. \ref{eq:rejuvenation}. The majority of merger remnants are rejuvenated by a few Myr. In the fiducial model, about 25\% of systems have age differences between 0.5–2 Myr, and another 25\% between 2–5.7 Myr, with the merger remnant appearing younger in both cases. 

However, we also find that up to 23\% of merger remnants appear older than their tertiary companions. These systems originate from mergers with highly unequal mass ratios ($q_{\rm{in}} < 0.25$). Due to mass loss during mass transfer and the merger itself, the final remnant is often less massive than the original primary, leading to an older apparent age. Conversely, the systems with the largest rejuvenation of the merger remnant, with age differences of several tens of Myr, originate from binaries that favour more equal mass components ($q_{\rm{in}} > 0.5$) and lower primary masses, which have longer evolutionary timescales.

Accurate age determination of massive MS binaries through observational measurements, such as the system reported by \citet{frost2024}, requires accurate determination of the system's properties via, e.g., spectroscopic and interferometric observations. For reference, the large sample of spectroscopic binaries in the Large Magellanic Cloud reported by \citet{mahy2020} have uncertainties ranging from about 10\% to over 50\% of the current star's ages. Across our model variations, an age uncertainty of 10\% for both stars would result in 73-83\% of the binaries able to be characterised as non-coeval just after the merger, while with an uncertainty of 50\%, this would reduce to 21-31\%. 

We do not find any significant difference in the age discrepancies across the evolutionary channels described in Sect. \ref{sect:post_merger}. Similarly, we do not predict any significant correlation between the degree of rejuvenation and the properties of the post-merger orbital configuration or tertiary mass. Nevertheless, as most post-merger binaries do not interact further, typical orbits of these non-coeval binaries are wide and eccentric. This is also underlined by \citet{shariat2025}, who investigated the formation of BSSs through mergers in hierarchical triples of low- to intermediate-mass stars. They predict that 26-54\% of wide ($a_{\rm{pm}} \gtrsim 100\,\rm{AU}$) double WD binaries host a merger product.

\subsubsection{When the lower-mass star evolves first}

We identify a subpopulation in which the tertiary star reaches the terminal-age MS (TAMS) before the merger remnant does, despite having a lower mass. From a perspective of isolated binary evolution this is counterintuitive, as the more massive companion is expected to evolve more rapidly. In our simulations, this behaviour occurs in approximately 5–14\% of systems. It is more prevalent when the mass transfer is conservative, and rejuvenation of the merger remnant is more prominent.

Most of these systems (75–81\%) evolve without further interaction. However, in 16–24\% of cases, the tertiary eventually fills its Roche lobe. As that happens, mass transfer is predominantly stable, as the donor is initially less massive than the accretor, except in systems where the donor has a deep convective envelope.
When the mass transfer is sufficiently conservative, the orbit widens steadily throughout the interaction. 

As discussed in Sect. \ref{sec:mt_to_merger}, systems with mass transfer from the original tertiary to the merger remnant tend to evolve toward longer post-interaction orbital periods than those in which the merger remnant is the donor. This difference is largely due to the fact that most tertiary donors have a mass that is lower than the mass of the merger remnant. To illustrate the evolution of such system, we showcase an example system from our simulated population. Initially, the stellar masses are $m_1=9.0 \, \rm{M}_{\odot}$, $m_2=4.2 \, \rm{M}_{\odot}$, and $m_3=10.1 \, \rm{M}_{\odot}$, with inner and outer orbital separations of $12.4 
 \, \rm{R}_{\odot}$ and $2752 \, \rm{R}_{\odot}$, and eccentricities of 0.17 and 0.93, respectively. After 17.6 Myr, the primary star (still on the MS) fills its Roche lobe, followed by a short period of stable, quite non-conservative mass transfer on a thermal timescale. This results in a stellar merger that produces a rejuvenated MS star of $11.4 \, \rm{M}_{\odot}$. As a result of mass loss, the outer orbit widens to $2996 \, \rm{R}_{\odot}$. At 24 Myr, the original tertiary enters the HG and fills its Roche lobe shortly afterward. By this point, the orbit has circularised at a separation of $402 \, \rm{R}_{\odot}$. The tertiary, throughout its core-helium burning phase, continues transferring its envelope over a span of 0.3 Myr. The system ends up as a stripped tertiary of $2.3 \, \rm{M}_{\odot}$, and the merger remnant of $19.0 \, \rm{M}_{\odot}$. The resulting orbital separation is $2569 \, \rm{R}_{\odot}$. Eventually, the tertiary forms a NS, imparting a SN kick that alters the orbit to a separation of $4963 \, \rm{R}_{\odot}$ and an eccentricity of 0.93. After 29.5 Myr, as the merger remnant becomes a supergiant, the system undergoes a Darwin-Riemann instability \citep{hut1980}, leading to rapid inspiral of the binary and a final separation of $7.7 \, \rm{R}_{\odot}$. After 30.8 Myr, the merger remnant also forms a NS. The resulting system has a separation of $16.8 \, \rm{R}_{\odot}$ and an eccentricity of 0.65.

\subsubsection{Compact objects}

In Sect. \ref{sect:post_merger}, we described the population of post-merger systems up to the end of mass transfer between the merger remnant and the tertiary companion. In this section, we will extend our analysis toward later evolutionary phases, specifically to the point where the stars have evolved into a CO.

We start by considering systems in which only one of the stars has evolved into a CO. Only a fraction of the binaries (27-41\%) remain gravitationally bound during the preceding SN event. Disruptions easily occur in systems with wide pre-SN orbits, hence modest gravitational binding, when the imparted kick at NS formation is sizeable. Among the surviving systems, over half are wide, non-interacting binaries with orbital periods exceeding $10^4$ days. 

The binaries with shorter periods have experienced a prior phase of mass transfer (see Sect. \ref{sect:post_merger}), and may become detectable as high-mass X-ray binaries (HMXBs) via either a subsequent phase of mass transfer or wind accretion onto the CO, provided the CO accretes a sufficient amount of material. Traditionally, the orbits that lead to the formation of HMXBs are thought to result from an earlier phase of stable, conservative mass transfer in initially close binaries. In our models, post-merger orbits are inherently wide, and cannot produce the tight orbits required for efficient X-ray emission via such pathway. Alternative channels involving non-conservative or CE evolution have also been proposed, particularly to explain the properties of Be/X-ray binaries \citep{pfahl2002,podsiadlowski2004} and short-period supergiant HMXBs \citep{tauris2023}. Also, previous studies suggest that triple evolution is crucial for the formation of low-mass X-ray binaries \citep{podsiadlowski2003,naoz_xray2016,shariat_xray2025} Here, we investigate whether post-merger triples are able to form potentially detectable HMXBs. We also explore whether the components of the binary might show signatures that betray its triple origin. We examine this by determining the contribution of both post-merger mass transfer channels --- mass transfer from the merger remnant to the original tertiary and vice versa --- to the HMXB population. We note that in systems where the merger remnant is the CO, it may be hard to deduce that the system experienced a prior stellar merger. This is likely easier to do if the merger remnant is still a star.

Throughout this analysis, we only consider masses of the stellar companion of at least $8 \, \rm{M}_{\odot}$. The detectability of X-ray binaries scales with their X-ray luminosity, which depends on the accretion rate onto the CO. We estimate the capture efficiency of material by the CO using the Bondi-Hoyle-Lyttleton formalism. For circular orbits, the accretion efficiency $\eta$ is given by:
\begin{equation}
    \eta = \Bigg[\frac{\tilde{q}}{(1+v_w/v_{\rm{circ}})^2}\Bigg]^2,
\end{equation}
where $\tilde{q} = m_2/(m_1+m_2)$ is the mass ratio with $m_1$ the mass of the stellar companion and $m_2$ the mass of the CO, $v_w$ the wind velocity of the stellar companion, and $v_{\rm{circ}}$ the orbital velocity \citep{tejeda2025}. We assume the wind velocity to be constant and $v_w \gg v_{\rm{circ}}$, such that $\eta \propto \tilde{q}^2 / v_{\rm{circ}}^4$. This assumption is justified given the typically high terminal wind velocities of OB stars \citep{vink2021}.

\begin{figure}
\includegraphics[width=\hsize]{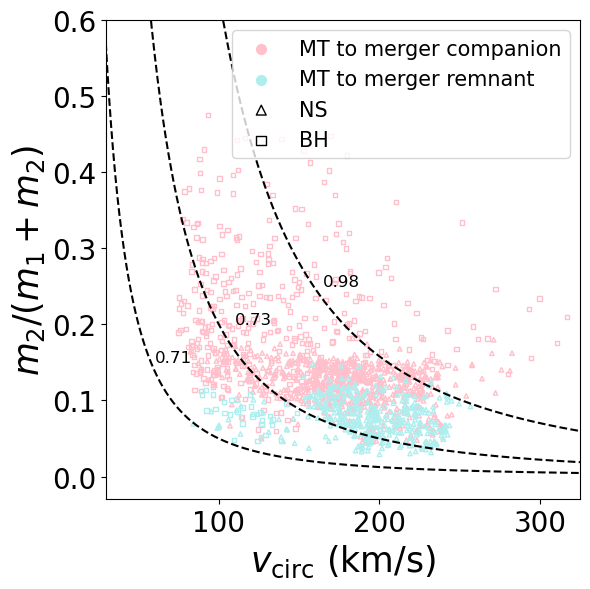}
    \caption{Circular orbital velocity  and mass ratio $\tilde{q} = m_2/(m_1+m_2)$ for binaries containing a CO and post-MS high-mass companion in the fiducial model. \textit{Left:} We differentiate between systems where the post-merger remnant is the CO (pink), and where the post-merger remnant is the stellar companion (blue). The dashed black curves indicate lines of constant accretion efficiency proxy $\eta \propto \tilde{q}^2 / v_{\rm{circ}}^4$, with labelled values showing the fraction of systems above each line where the CO is the post-merger remnant. We distinguish the systems with a NS companion (triangles) and a BH companion (squares).}
    \label{fig:figure8}
\end{figure}

Our simulations predict that between 4.7-12.0\% of all mergers form a binary consisting of a CO and an evolved stellar companion with a period less than a year, corresponding to the period range of observed HMXBs \citep{fortin2023}. In most of these systems (65-93\%), it is the merger remnant that has evolved into a CO, as shown in Fig. \ref{fig:figure8}. Toward higher accretion efficiencies and shorter periods (a few to a few tens of days), which forms the bulk of the observed HMXB population, their contribution is the highest. HMXBs where the original tertiary is the CO are less likely to form, due to a combination of wider post-mass-transfer orbits, more unequal mass ratios (which reduce the accretion efficiency), and the overall lower frequency of systems where the tertiary fills its Roche lobe before the merger remnant does.

The properties of the HMXBs differ depending on which star initiates mass transfer after the inner binary has merged. In the fiducial model, if the merger remnant fills its Roche lobe first, we find that 65\% of the COs are a NS, with stellar companions predominantly near the lower mass threshold of $8 \, \rm{M_{\odot}}$. In contrast, if the original tertiary fills its Roche lobe first, the system forms a NSs in 92\% of cases. Additionally, their companions are generally more massive, peaking around $15 \, \rm{M}_{\odot}$, leading to more unequal mass ratios. Taken together, we predict that in our triple channel about three-quarters of the HMXBs formed contain a NS rather than a BH, which is fewer compared to the observed ratio of NS to BH systems in the HMXB population \citep{neumann2023}.

Binaries with intermediate periods (100-1000 days) that consist of a non-interacting MS-star and a CO companion may represent an interesting population detectable with \textit{Gaia} \citep{el-badry_sun2023,el-badry_giant2023,el-badry2024,gaia2024}. The formation of some of these observed systems has been linked to stellar mergers within triples \citep{generozov2024,li2024,li2025,regaly2025}. We predict that a non-negligible fraction of mergers (2.7-6.0\%) produce MS+CO binaries in this period range, with the slight majority containing a NS. Systems with more massive BHs typically undergo prior mass transfer onto the MS companion, which limits the maximum BH mass in our simulations to about $10\,\rm{M}_{\odot}$. However, \cite{li2024} shows that such mass transfer can be avoided, as stellar merger remnants may retain smaller radii throughout their evolution compared to single stars of the same mass, an effect that is not included in our models (see also Sect. \ref{sec:bsg}).

Within the simulated population, 5-18\% of all systems with a stellar merger remain bound after the formation of a second CO. If the orbit of the resulting DCOs are compact enough, they may ultimately merge and become observable through GW emission. About half of these binaries form wide BBHs that did not undergo any prior interaction with their companion. Such wide BBHs can merge by increasing their eccentricity through torques exerted by the Galactic potential \citep{stegmann2024}. The DCOs that have experienced mass transfer after the stellar merger have distinct characteristics in their orbital period distribution. A prominent peak between 1–100 days is associated with systems that underwent CE evolution during the second mass-transfer phase; these primarily evolve into BH+BH binaries. A secondary peak at much shorter periods, around 0.1 days, consists of mainly NS+NS binaries. The latter experienced an additional phase of mass transfer from a (nearly) stripped donor to the CO, further stripping the donor and decreasing the orbital separation.

We estimate GW inspiral times using the analytical prescription from \citet{mandel2021}, and find that only systems with the shortest periods, primarily NS+NS binaries and a smaller fraction of BH+NS binaries, are able to merge within a Hubble time. Overall, we predict that 0.9–6.6\% of massive triple systems that underwent a stellar merger will ultimately produce GW mergers. We emphasise that these results are based on solar metallicity simulations, and DCO mergers are known to be more efficiently produced at lower metallicities. \citet{stegmann2022b} show that at low metallicity, tertiary stars which are more evolved but less massive than the merger remnant can produce BH+BH mergers with more unequal mass ratios than expected from isolated binary evolution, highlighting the potential of stellar mergers in triples to diversify the GW merger population. In addition to the stellar-merger pathway explored here, prior studies have shown that hierarchical triples in the field can also drive DCO mergers through secular perturbations from the tertiary star, leading to observable signatures in the eccentricity distribution of mergers \citep{antonini2017,silsbee2017S,fragione2019,bartos2023,dorozsmai2024,vigna-gomez2025,stegmann2025}.

\section{Discussion}
\label{sec:sec4}
\subsection{Galactic rates}
We estimate Galactic merger rates based on our simulated populations, assuming a constant Galactic star-formation rate (SFR) of $2 \, \rm{M}_{\odot}\,\rm{yr}^{-1}$ \citep{chomiuk2011,elia2022}. We define the merger rate as:
\begin{equation}
    \mathcal{R}_{\rm{merger}} = \mathcal{R}_{\rm{birth}}\times f_{\rm{sim}}\times f_{\rm{merge}},
\end{equation}
where $\mathcal{R}_{\rm{birth}}$ is the average number of stellar systems born per year, $f_{\rm{sim}}$ the fraction of the initial stellar and orbital parameter space sampled by the simulations compared to the full range expected in a real stellar population, and $f_{\rm{merge}}$ the fraction of simulated triples that undergo a stellar merger. The birthrate is calculated as the SFR divided by the average stellar system mass $\tilde{M}$. We derive $\tilde{M}$ assuming a triple fraction of 0.57, binary fraction of 0.36, and single-star fraction of 0.07 \citep{moe2017}, with stellar masses between $0.08$ and $150\,\rm{M}_\odot$. Here, we neglect higher-order multiples. Since we simulated only triple systems with massive primaries, the fraction of sampled systems is given by $f_{\rm{sim}} = f_{\rm{triple}}\times f_{\rm{M_{1,ZAMS}}}$, where $f_{\rm{triple}}$ is the triple fraction and $f_{\rm{M_{1,ZAMS}}}$ the fraction of stars with ZAMS masses between $8$ and $100\,\rm{M}_\odot$. For the fiducial model, we obtain $\mathcal{R}_{\rm{birth}} = 1.8\,\rm{yr}^{-1}$, $f_{\rm{sim}} = 3.6\times10^{-3}$, and $f_{\rm{merge}} = 0.29$.

We find a Galactic merger rate of $(1.3-2.2)\times10^{-3} \, \rm{yr}^{-1}$ for massive triple systems, depending on the model assumptions. The corresponding values for all model variations are listed in Table \ref{table:merger_fraction}. This rate is comparable to those predicted for massive binary mergers from binary population synthesis \citep[e.g.,][]{kochanek2014}. Despite the low predicted rate, a few extra-galactic massive merger candidates have been detected, identified as luminous red novae \citep{pastorello2019}.

We also estimate the fraction of wide MS+MS binaries in the Galaxy that contain a merger remnant. To this end, we ran a comparison population of isolated binaries with initial separations $>100\,\rm{R}_{\odot}$, ensuring the systems experience no mass transfer. This threshold corresponds approximately to the minimum separation found for post-merger binaries originating from triples.  To assess the relative contribution of triple versus binary evolution, we consider three factors. First, The initial multiplicity fractions of massive stars, with $f_{\rm{triple}} = 0.57$ and $f_{\rm{binary}} = 0.36$. Second, the fraction of systems that form a MS+MS binary with a separation above $100 \, \rm{R}_{\odot}$ and at least one component above $8\,\rm{M}_{\odot}$. For the binary population, we assume a distribution of initial orbital separations that is uniform in logarithmic space. Third, the time such systems spend in the MS+MS phase, which affects the probability of observing them at a given time.

Combining our binary and triple simulations, we predict that 13-15\% of wide MS+MS binaries have a component that experienced a past stellar merger, when we assume a constant SFR over at least the past 100 Myr. The contribution from isolated binaries is relatively robust, as it is not sensitive to stellar interaction physics, but it does depend on the assumed initial separation distribution. If the true distribution favours closer orbits \citep{sana2012}, the contribution from isolated binaries would be lower. Furthermore, our models exclude primaries with initial masses below $8\,\rm{M}_{\odot}$, which could still form massive stars through mergers \citep[e.g.,][]{zapartas2017} and hence increase the fraction of MS+MS binaries having a merger history.

If we assume that the ages of both stellar components have measuring uncertainties of 10\% (50\%), we find that 8-10\% (2-3\%) of the wide MS+MS binaries would be identified as non-coeval. Furthermore, when focussing on systems with orbital periods $\lesssim10^4$ days, roughly within the current observational capabilities of spectroscopic and interferometric methods \citep[e.g.,][]{sana2025}, the observed fraction of non-coeval binaries is 6-7\% ($\sim$2\%).

We note that the percentages we quote reflect only the contribution from non-coeval binaries formed through stellar mergers in triples, relative to a reference population of isolated binaries. In practice, hierarchical triples can produce wide MS+MS binaries via alternative pathways. The most important is when the tertiary is initially the most massive star and the system becomes unbound following a SN kick, leaving behind a wide MS+MS binary composed of the inner binary. In our simulations, this pathway accounts for up to 10\% of all triples and is therefore non-negligible. However, the occurrence rate is highly sensitive to assumptions about the tertiary mass distribution and the SN kick magnitude. We do not vary assumptions regarding the SN kick, but a lower kick velocity would increase the number of bound systems, thereby reducing this channel’s contribution. Other potential formation pathways for wide MS+MS binaries in triples are not expected to impact the overall statistics much. 

\subsection{Blue supergiants}
\label{sec:bsg}
A crucial limitation of using stellar population synthesis for multiple star systems, is that the internal evolution of the star is based on single-star models. In reality, stellar interactions can alter the internal structure of the donor and accretor stars \citep[e.g.,][]{podsiadlowski2004,laplace2021,schneider2021,schneider2024}, affecting their subsequent evolution. Recent detailed simulations show that post-MS accretors, including merger remnants, can evolve as blue supergiants (BSGs), rather than transitioning into red supergiants (RSGs) \citep{schneider2024}. 

Whether a star remains a BSG or eventually moves to the RSG branch depends primarily on the fraction of mass it accretes during the interaction. \citet{schneider2024} show that stars shortly after completing core hydrogen burning can end their lives as BSGs if they accrete more than 75\% of their mass. BSGs experience lower mass-loss rates compared to RSGs, leading to higher final masses. More relevant for this study, BSGs also attain smaller maximum radii compared to single stars of similar mass. If a merger remnant stays in the BSG phase prior to core collapse, it is unlikely to initiate mass transfer with the tertiary.

In our models, 10\% of the systems that undergo mass transfer from the merger remnant to the original tertiary had a post-MS donor at the time of merger, typically early on the HG. In the fiducial model, about one-fifth of these pre-merger systems have an inner mass ratio $q_{\rm{in}}>0.75$, indicating that, if the merger is conservative, the remnant may accrete enough mass to evolve into and end its life as a BSG. Although our population synthesis models do not explicitly follow BSG evolution, this limitation is unlikely to significantly affect our global predictions. The subsequent evolution of these remnants could give rise to particularly interesting systems.

\subsection{Contact binaries}
The evolution and fate of contact binaries is still poorly understood. Observational evidence suggests that some contact systems are long-lived \citep{abdul-masih2021,menon2024}, but theoretical models indicate that their stability depends on the initial orbital period and mass ratio \citep{menon2021,henneco2024}. In particular, \citet{henneco2024} find that massive binaries with initial mass ratios below 0.5, undergoing case A mass transfer, are expected to merge on a thermal timescale shortly after entering contact. For such systems, our simplified assumption of an immediate merger appears justified. Conversely, for contact systems with more equal mass ratios, our approach is less certain. They may actually survive for an extended period, in line with the observed population of contact binaries. However, current theoretical models still struggle to reproduce observable properties such as luminosities and mass ratios \citep{fabry2025}. This is particularly relevant, as the outcome of the contact phase can span a wide range of post-contact products \citep{abdul-masih2025}.

In our simulated population, mass transfer between two MS stars is the most dominant interaction channel within the inner binary, also for the systems that merge. Of these systems,  70\% have an initial inner mass ratio $q_{\rm{in}}<0.5$. The distribution of initial mass ratios varies across different post-merger evolutionary pathways. Notably, in systems where the tertiary star later transfers mass to the merger remnant, $q_{\rm{in}}$ peaks just below 0.4, whereas other channels have more extreme mass ratios (see Fig. \ref{fig:figure1}). Despite these differences in distribution, the fraction of systems with $q_{\rm{in}} < 0.5$ remains comparable across channels. 

For systems with more equal mass ratios, the assumption of an immediate merger is less suitable. This likely leads to an overestimate of early main-sequence mergers among more equal-mass binaries. These binaries are also the ones most strongly rejuvenated, due to the higher mass accretion during the merger. While not dominant, these systems are non-negligible. A more sophisticated prescription for the contact phase, especially one that accounts for its duration, would improve the accuracy of merger rate predictions and the post-merger properties of the resulting stars.

\subsection{Simplifications in the pre-merger evolution}

In our modelling approach, we transition from simulating the system with the triple evolution code {\tt TRES} to the binary evolution code {\tt SeBa} once the first mass-transfer phase begins, despite the system still being a hierarchical triple. Hence, dynamical interactions within the system are neglected once the switch to {\tt SeBa} is made. This is an oversimplification for systems in which dynamical perturbations enhance the eccentricity of the inner binary, and thus the likelihood of mass transfer. Depending on the duration of the mass transfer, continued three-body dynamics could modify the inner orbit, possibly altering the nature or timing of mergers and collisions. We expect such effects to be most important in systems with long-lived, stable mass transfer, such as RLOF occurring on the nuclear timescale of the donor. \citet{hamers2019} demonstrated through an example that when the mass transfer rate is low, the inner binary can experience many ZLK oscillations before becoming dynamically decoupled from the tertiary companion. However, they did not include the effect of tidal forces, which would mitigate the three-body dynamical effects. In contrast, for short-lived mass phases such as common-envelope evolution or mass transfer on the thermal-timescale, this approximation is reasonable. 

To quantify this, we find that among systems where dynamical interactions increased the inner eccentricity with at least 0.05, accounting for 12.5\% of all mergers, approximately 40\% undergo mass transfer that lasts at least as long as the evolution preceding it. Assessing the full impact on the mass transfer would require a self-consistent treatment that couples mass transfer, tidal evolution, and three-body dynamical interactions.

Additionally, we assume the inner orbit is circularised at the onset of mass transfer. This is a standard assumption in binary evolution codes, as tidal forces are generally expected to circularise the orbit prior to RLOF. However, in hierarchical triples, secular perturbations can prevent the orbit from circularising upon RLOF \citep{toonen2020,kummer2023}. While this may not substantially alter the overall merger rate, it could affect the nature of the merger event itself. For example, an eccentric orbit at the time of merger could result in more direct, head-on collisions \citep[e.g.,][]{hamers2019}.

\section{Conclusion}
\label{sec:sec5}
In this study, we investigated the incidence rate and properties of stellar mergers and their post-merger evolution within massive hierarchical triples in the Galaxy. In this regard, we performed rapid population synthesis simulations of massive triple stars at solar metallicity. We included a set of model variations to probe the impact of  uncertainties in the initial properties of the triples, mass transfer physics, and merger physics.

A notable fraction ($\sim$20-30\%) of the triples experiences a stellar merger as a result of a mass-transfer phase. Due to initially compact inner orbits, and potential strong three-body interactions, mergers generally take place early in the evolution of the primary star; over 60\% of mergers occur between two MS stars, and most of the remaining mergers have a donor early on the HG (see Fig. \ref{fig:figure2}). The MS+MS mergers are expected to form a more massive, rejuvenated MS star, whose merger past could be characterised by the apparent age discrepancy with the MS tertiary companion (see Fig. \ref{fig:figure7}). We estimate that age discrepancies can be observationally detected in 73-83\% (21-31\%) of systems when the age uncertainty is 10\% (50\%). These results are most sensitive to assumptions about the conservativeness of mass transfer. Predicted merger rates may deviate up to 30\% relative to our fiducial model. The efficiency of the CE phase and the initial mass of the tertiary star also noticeably affect the predicted outcomes, though to a lesser extent.

The post-merger orbits cover a large range of periods, mass ratios and eccentricities. However, there is a clear preference toward wide, eccentric orbits (see Fig. \ref{fig:figure4}). Consequently, around 80\% of the post-merger remnants will have no further interactions with their tertiary companion. Additionally, there is a dearth of short-period orbits directly after merger (less than 100 days), which arises from the dynamical stability requirements inherent to hierarchical triple systems. Among post-merger binaries that are sufficiently compact to engage in mass transfer, either the merger remnant or the tertiary can be the first to fill their Roche lobe as a post-MS donor, although the merger remnant significantly more often. Due to the rejuvenated nature of MS+MS merger remnants, the star can leave the MS later than its companion, even if it is higher in mass. This results in the mass transfer being primarily stable for tertiary donors (see Fig. \ref{fig:figure6}) and their post-mass transfer orbits being exceptionally wide compared to what would be expected for standard binary evolution.

Subsequent evolution can result in short-period compact-object binaries, such as HMXBs (see Fig. \ref{fig:figure8}) or GW sources. Since the merger remnant typically evolves first into a NS or BH, any direct signatures of its merger history, and thus the triple origin, may be lost, complicating observational identification.

We estimate a Galactic stellar merger rate from massive triples of $(1.3-2.2)\times10^{-3} \, \rm{yr}^{-1}$. Although this is relatively low, a non-negligible fraction (2-10\%) of wide ($P \gtrsim 100$ days) MS+MS binaries are predicted to have detectable age discrepancies due to a past merger that may serve as a way to identify merged stars.

\begin{acknowledgements}
    FK would like to thank Eva Laplace, Jakob van den Eijnden, Julia Bodensteiner, Tomer Shenar, and Zsolt Keszthelyi for the helpful discussions. The authors also acknowledge support from the Netherlands Research Council NWO (VIDI 203.061 grant).
\end{acknowledgements}

  \bibliographystyle{aa}
  \bibliography{refs.bib}

\begin{thebibliography}{143}
\expandafter\ifx\csname natexlab\endcsname\relax\def\natexlab#1{#1}\fi

\bibitem[{{Abdul-Masih}(2025)}]{abdul-masih2025}
{Abdul-Masih}, M. 2025, Contributions of the Astronomical Observatory Skalnate
  Pleso, 55, 390

\bibitem[{{Abdul-Masih} {et~al.}(2021){Abdul-Masih}, {Sana}, {Hawcroft},
  {Almeida}, {Brands}, {de Mink}, {Justham}, {Langer}, {Mahy}, {Marchant},
  {Menon}, {Puls}, \& {Sundqvist}}]{abdul-masih2021}
{Abdul-Masih}, M., {Sana}, H., {Hawcroft}, C., {et~al.} 2021, \aap, 651, A96

\bibitem[{{Antonini} {et~al.}(2017){Antonini}, {Toonen}, \&
  {Hamers}}]{antonini2017}
{Antonini}, F., {Toonen}, S., \& {Hamers}, A.~S. 2017, \apj, 841, 77

\bibitem[{{Bartos} {et~al.}(2023){Bartos}, {Rosswog}, {Gayathri}, {Miller},
  {Veske}, \& {Marka}}]{bartos2023}
{Bartos}, I., {Rosswog}, S., {Gayathri}, V., {et~al.} 2023, arXiv e-prints,
  arXiv:2302.10350

\bibitem[{{Bashi} \& {Tokovinin}(2024)}]{bashi2024}
{Bashi}, D. \& {Tokovinin}, A. 2024, \aap, 692, A247

\bibitem[{{Bavera} {et~al.}(2023){Bavera}, {Fragos}, {Zapartas}, {Andrews},
  {Kalogera}, {Berry}, {Kruckow}, {Dotter}, {Kovlakas}, {Misra}, {Rocha},
  {Srivastava}, {Sun}, \& {Xing}}]{bavera2023}
{Bavera}, S.~S., {Fragos}, T., {Zapartas}, E., {et~al.} 2023, Nature Astronomy,
  7, 1090

\bibitem[{{Belczynski} {et~al.}(2010){Belczynski}, {Bulik}, {Fryer}, {Ruiter},
  {Valsecchi}, {Vink}, \& {Hurley}}]{belczynski2010}
{Belczynski}, K., {Bulik}, T., {Fryer}, C.~L., {et~al.} 2010, \apj, 714, 1217

\bibitem[{{Bj{\"o}rklund} {et~al.}(2021){Bj{\"o}rklund}, {Sundqvist}, {Puls},
  \& {Najarro}}]{bjorklund2021}
{Bj{\"o}rklund}, R., {Sundqvist}, J.~O., {Puls}, J., \& {Najarro}, F. 2021,
  \aap, 648, A36

\bibitem[{{Borkovits} {et~al.}(2015){Borkovits}, {Rappaport}, {Hajdu}, \&
  {Sztakovics}}]{borkovits2015}
{Borkovits}, T., {Rappaport}, S., {Hajdu}, T., \& {Sztakovics}, J. 2015,
  \mnras, 448, 946

\bibitem[{{Broekgaarden} {et~al.}(2021){Broekgaarden}, {Berger}, {Neijssel},
  {Vigna-G{\'o}mez}, {Chattopadhyay}, {Stevenson}, {Chruslinska}, {Justham},
  {de Mink}, \& {Mandel}}]{broekgaarden2021}
{Broekgaarden}, F.~S., {Berger}, E., {Neijssel}, C.~J., {et~al.} 2021, \mnras,
  508, 5028

\bibitem[{{Broekgaarden} {et~al.}(2022){Broekgaarden}, {Berger}, {Stevenson},
  {Justham}, {Mandel}, {Chru{\'s}li{\'n}ska}, {van Son}, {Wagg},
  {Vigna-G{\'o}mez}, {de Mink}, {Chattopadhyay}, \&
  {Neijssel}}]{broekgaarden2022}
{Broekgaarden}, F.~S., {Berger}, E., {Stevenson}, S., {et~al.} 2022, \mnras,
  516, 5737

\bibitem[{{Burdge} {et~al.}(2024){Burdge}, {El-Badry}, {Kara}, {Canizares},
  {Chakrabarty}, {Frebel}, {Millholland}, {Rappaport}, {Simcoe}, \&
  {Vanderburg}}]{burdge2024}
{Burdge}, K.~B., {El-Badry}, K., {Kara}, E., {et~al.} 2024, \nat, 635, 316

\bibitem[{{Camacho} {et~al.}(2014){Camacho}, {Torres}, {Garc{\'\i}a-Berro},
  {Zorotovic}, {Schreiber}, {Rebassa-Mansergas}, {Nebot G{\'o}mez-Mor{\'a}n},
  \& {G{\"a}nsicke}}]{camacho2014}
{Camacho}, J., {Torres}, S., {Garc{\'\i}a-Berro}, E., {et~al.} 2014, \aap, 566,
  A86

\bibitem[{{Chen} \& {Han}(2008)}]{chen2008}
{Chen}, X. \& {Han}, Z. 2008, \mnras, 387, 1416

\bibitem[{{Chomiuk} \& {Povich}(2011)}]{chomiuk2011}
{Chomiuk}, L. \& {Povich}, M.~S. 2011, \aj, 142, 197

\bibitem[{{de Kool}(1990)}]{dekool1990}
{de Kool}, M. 1990, \apj, 358, 189

\bibitem[{{de Mink} {et~al.}(2014){de Mink}, {Sana}, {Langer}, {Izzard}, \&
  {Schneider}}]{demink2014}
{de Mink}, S.~E., {Sana}, H., {Langer}, N., {Izzard}, R.~G., \& {Schneider},
  F.~R.~N. 2014, \apj, 782, 7

\bibitem[{{Dorozsmai} {et~al.}(2024){Dorozsmai}, {Toonen}, {Vigna-G{\'o}mez},
  {de Mink}, \& {Kummer}}]{dorozsmai2024}
{Dorozsmai}, A., {Toonen}, S., {Vigna-G{\'o}mez}, A., {de Mink}, S.~E., \&
  {Kummer}, F. 2024, \mnras, 527, 9782

\bibitem[{{El-Badry} {et~al.}(2023{\natexlab{a}}){El-Badry}, {Rix}, {Cendes},
  {Rodriguez}, {Conroy}, {Quataert}, {Hawkins}, {Zari}, {Hobson}, {Breivik},
  {Rau}, {Berger}, {Shahaf}, {Seeburger}, {Burdge}, {Latham}, {Buchhave},
  {Bieryla}, {Bashi}, {Mazeh}, \& {Faigler}}]{el-badry_giant2023}
{El-Badry}, K., {Rix}, H.-W., {Cendes}, Y., {et~al.} 2023{\natexlab{a}},
  \mnras, 521, 4323

\bibitem[{{El-Badry} {et~al.}(2024){El-Badry}, {Rix}, {Latham}, {Shahaf},
  {Mazeh}, {Bieryla}, {Buchhave}, {Andrae}, {Yamaguchi}, {Isaacson}, {Howard},
  {Savino}, \& {Ilyin}}]{el-badry2024}
{El-Badry}, K., {Rix}, H.-W., {Latham}, D.~W., {et~al.} 2024, The Open Journal
  of Astrophysics, 7, 58

\bibitem[{{El-Badry} {et~al.}(2023{\natexlab{b}}){El-Badry}, {Rix}, {Quataert},
  {Howard}, {Isaacson}, {Fuller}, {Hawkins}, {Breivik}, {Wong}, {Rodriguez},
  {Conroy}, {Shahaf}, {Mazeh}, {Arenou}, {Burdge}, {Bashi}, {Faigler}, {Weisz},
  {Seeburger}, {Almada Monter}, \& {Wojno}}]{el-badry_sun2023}
{El-Badry}, K., {Rix}, H.-W., {Quataert}, E., {et~al.} 2023{\natexlab{b}},
  \mnras, 518, 1057

\bibitem[{{Elia} {et~al.}(2022){Elia}, {Molinari}, {Schisano}, {Soler},
  {Merello}, {Russeil}, {Veneziani}, {Zavagno}, {Noriega-Crespo}, {Olmi},
  {Benedettini}, {Hennebelle}, {Klessen}, {Leurini}, {Paladini}, {Pezzuto},
  {Traficante}, {Eden}, {Martin}, {Sormani}, {Coletta}, {Colman}, {Plume},
  {Maruccia}, {Mininni}, \& {Liu}}]{elia2022}
{Elia}, D., {Molinari}, S., {Schisano}, E., {et~al.} 2022, \apj, 941, 162

\bibitem[{{Fabry} {et~al.}(2025){Fabry}, {Marchant}, {Langer}, \&
  {Sana}}]{fabry2025}
{Fabry}, M., {Marchant}, P., {Langer}, N., \& {Sana}, H. 2025, \aap, 695, A109

\bibitem[{{Fortin} {et~al.}(2023){Fortin}, {Garc{\'\i}a}, {Simaz Bunzel}, \&
  {Chaty}}]{fortin2023}
{Fortin}, F., {Garc{\'\i}a}, F., {Simaz Bunzel}, A., \& {Chaty}, S. 2023, \aap,
  671, A149

\bibitem[{{Fragione} \& {Loeb}(2019)}]{fragione2019}
{Fragione}, G. \& {Loeb}, A. 2019, \mnras, 486, 4443

\bibitem[{{Frost} {et~al.}(2025){Frost}, {Sana}, {Le Bouquin}, {Perets},
  {Bodensteiner}, {Igoshev}, {Banyard}, {Mahy}, {M{\'e}rand}, \&
  {Ram{\'\i}rez-Agudelo}}]{frost2025}
{Frost}, A.~J., {Sana}, H., {Le Bouquin}, J.~B., {et~al.} 2025, \aap, 701, A171

\bibitem[{{Frost} {et~al.}(2024){Frost}, {Sana}, {Mahy}, {Wade}, {Barron}, {Le
  Bouquin}, {M{\'e}rand}, {Schneider}, {Shenar}, {Barb{\'a}}, {Bowman},
  {Fabry}, {Farhang}, {Marchant}, {Morrell}, \& {Smoker}}]{frost2024}
{Frost}, A.~J., {Sana}, H., {Mahy}, L., {et~al.} 2024, Science, 384, 214

\bibitem[{{Fryer} {et~al.}(2012){Fryer}, {Belczynski}, {Wiktorowicz},
  {Dominik}, {Kalogera}, \& {Holz}}]{fryer2012}
{Fryer}, C.~L., {Belczynski}, K., {Wiktorowicz}, G., {et~al.} 2012, \apj, 749,
  91

\bibitem[{{Gaia Collaboration} {et~al.}(2024){Gaia Collaboration}, {Panuzzo},
  {Mazeh}, {Arenou}, {Holl}, {Caffau}, {Jorissen}, {Babusiaux}, {Gavras},
  {Sahlmann}, {Bastian}, {Wyrzykowski}, {Eyer}, {Leclerc}, {Bauchet},
  {Bombrun}, {Mowlavi}, {Seabroke}, {Teyssier}, {Balbinot}, {Helmi}, {Brown},
  {Vallenari}, {Prusti}, {de Bruijne}, {Barbier}, {Biermann}, {Creevey},
  {Ducourant}, {Evans}, {Guerra}, {Hutton}, {Jordi}, {Klioner}, {Lammers},
  {Lindegren}, {Luri}, {Mignard}, {Nicolas}, {Randich}, {Sartoretti},
  {Smiljanic}, {Tanga}, {Walton}, {Aerts}, {Bailer-Jones}, {Cropper},
  {Drimmel}, {Jansen}, {Katz}, {Lattanzi}, {Soubiran}, {Th{\'e}venin}, {van
  Leeuwen}, {Andrae}, {Audard}, {Bakker}, {Blomme}, {Casta{\~n}eda}, {De
  Angeli}, {Fabricius}, {Fouesneau}, {Fr{\'e}mat}, {Galluccio}, {Guerrier},
  {Heiter}, {Masana}, {Messineo}, {Nienartowicz}, {Pailler}, {Riclet}, {Roux},
  {Sordo}, {Gracia-Abril}, {Portell}, {Altmann}, {Benson}, {Berthier},
  {Burgess}, {Busonero}, {Busso}, {Cacciari}, {C{\'a}novas}, {Carrasco},
  {Carry}, {Cellino}, {Cheek}, {Clementini}, {Damerdji}, {Davidson}, {de
  Teodoro}, {Delchambre}, {Dell'Oro}, {Fraile Garcia}, {Garabato},
  {Garc{\'\i}a-Lario}, {Haigron}, {Hambly}, {Harrison}, {Hatzidimitriou},
  {Hern{\'a}ndez}, {Hestroffer}, {Hodgkin}, {Jamal}, {Jevardat de Fombelle},
  {Jordan}, {Krone-Martins}, {Lanzafame}, {L{\"o}ffler}, {Lorca}, {Marchal},
  {Marrese}, {Moitinho}, {Muinonen}, {Nu{\~n}ez Campos}, {Oreshina-Slezak},
  {Osborne}, {Pancino}, {Pauwels}, {Recio-Blanco}, {Riello}, {Rimoldini},
  {Robin}, {Roegiers}, {Sarro}, {Schultheis}, {Smith}, {Sozzetti}, {Utrilla},
  {van Leeuwen}, {Weingrill}, {Abbas}, {{\'A}brah{\'a}m}, {Abreu Aramburu},
  {Ahmed}, {Altavilla}, {{\'A}lvarez}, {Anders}, {Anderson}, {Anglada Varela},
  {Antoja}, {Baig}, {Baines}, {Baker}, {Balaguer-N{\'u}{\~n}ez}, {Balog},
  {Barache}, {Barros}, {Barstow}, {Bartolom{\'e}}, {Bashi}, {Bassilana},
  {Baudeau}, {Becciani}, {Bedin}, {Bellas-Velidis}, {Bellazzini}, {Beordo},
  {Bernet}, {Bertolotto}, {Bertone}, {Bianchi}, {Binnenfeld},
  {Blanco-Cuaresma}, {Bland-Hawthorn}, {Blazere}, {Boch}, {Bossini},
  {Bouquillon}, {Bragaglia}, {Braine}, {Bratsolis}, {Breedt}, {Bressan},
  {Brouillet}, {Brugaletta}, {Bucciarelli}, {Butkevich}, {Buzzi}, {Camut},
  {Cancelliere}, {Cantat-Gaudin}, {Capilla Guilarte}, {Carballo}, {Carlucci},
  {Carnerero}, {Carretero}, {Carton}, {Casamiquela}, {Casey}, {Castellani},
  {Castro-Ginard}, {Ceraj}, {Cesare}, {Charlot}, {Chaudet}, {Chemin},
  {Chiavassa}, {Chornay}, \& {Chosson}}]{gaia2024}
{Gaia Collaboration}, {Panuzzo}, P., {Mazeh}, T., {et~al.} 2024, \aap, 686, L2

\bibitem[{{Generozov} \& {Perets}(2024)}]{generozov2024}
{Generozov}, A. \& {Perets}, H.~B. 2024, \apj, 964, 83

\bibitem[{{Glebbeek} {et~al.}(2013){Glebbeek}, {Gaburov}, {Portegies Zwart}, \&
  {Pols}}]{glebbeek2013}
{Glebbeek}, E., {Gaburov}, E., {Portegies Zwart}, S., \& {Pols}, O.~R. 2013,
  \mnras, 434, 3497

\bibitem[{{Glebbeek} \& {Pols}(2008)}]{glebbeek2008}
{Glebbeek}, E. \& {Pols}, O.~R. 2008, \aap, 488, 1017

\bibitem[{{Hamann} \& {Koesterke}(1998)}]{hamann1998}
{Hamann}, W.~R. \& {Koesterke}, L. 1998, \aap, 333, 251

\bibitem[{{Hamann} {et~al.}(1995){Hamann}, {Koesterke}, \&
  {Wessolowski}}]{hamann1995}
{Hamann}, W.~R., {Koesterke}, L., \& {Wessolowski}, U. 1995, \aap, 299, 151

\bibitem[{{Hamers} \& {Dosopoulou}(2019)}]{hamers2019}
{Hamers}, A.~S. \& {Dosopoulou}, F. 2019, \apj, 872, 119

\bibitem[{{Heintz} {et~al.}(2022){Heintz}, {Hermes}, {El-Badry}, {Walsh}, {van
  Saders}, {Fields}, \& {Koester}}]{heintz2022}
{Heintz}, T.~M., {Hermes}, J.~J., {El-Badry}, K., {et~al.} 2022, \apj, 934, 148

\bibitem[{{Heintz} {et~al.}(2024){Heintz}, {Hermes}, {Tremblay}, {Ould Rouis},
  {Reding}, {Kaiser}, \& {van Saders}}]{heintz2024}
{Heintz}, T.~M., {Hermes}, J.~J., {Tremblay}, P.~E., {et~al.} 2024, \apj, 969,
  68

\bibitem[{{Henneco} {et~al.}(2024){Henneco}, {Schneider}, \&
  {Laplace}}]{henneco2024}
{Henneco}, J., {Schneider}, F.~R.~N., \& {Laplace}, E. 2024, \aap, 682, A169

\bibitem[{{Hills}(1983)}]{hills1983}
{Hills}, J.~G. 1983, \apj, 267, 322

\bibitem[{{Hills} \& {Day}(1976)}]{hills1976}
{Hills}, J.~G. \& {Day}, C.~A. 1976, \aplett, 17, 87

\bibitem[{{Hubrig} {et~al.}(2008){Hubrig}, {Sch{\"o}ller}, {Schnerr},
  {Gonz{\'a}lez}, {Ignace}, \& {Henrichs}}]{hubrig2008}
{Hubrig}, S., {Sch{\"o}ller}, M., {Schnerr}, R.~S., {et~al.} 2008, \aap, 490,
  793

\bibitem[{{Hurley} {et~al.}(2000){Hurley}, {Pols}, \& {Tout}}]{hurley2000}
{Hurley}, J.~R., {Pols}, O.~R., \& {Tout}, C.~A. 2000, \mnras, 315, 543

\bibitem[{{Hurley} {et~al.}(2002){Hurley}, {Tout}, \& {Pols}}]{hurley2002}
{Hurley}, J.~R., {Tout}, C.~A., \& {Pols}, O.~R. 2002, \mnras, 329, 897

\bibitem[{{Hut}(1980)}]{hut1980}
{Hut}, P. 1980, \aap, 92, 167

\bibitem[{{Hut}(1981)}]{hut1981}
{Hut}, P. 1981, \aap, 99, 126

\bibitem[{{Hwang}(2023)}]{hwang2023}
{Hwang}, H.-C. 2023, \mnras, 518, 1750

\bibitem[{{Igoshev} {et~al.}(2020){Igoshev}, {Perets}, \&
  {Michaely}}]{igoshev2020}
{Igoshev}, A.~P., {Perets}, H.~B., \& {Michaely}, E. 2020, \mnras, 494, 1448

\bibitem[{{Ivanova}(2015)}]{ivanova2015}
{Ivanova}, N. 2015, in Astrophysics and Space Science Library, Vol. 413,
  Astrophysics and Space Science Library, ed. H.~M.~J. {Boffin}, G.~{Carraro},
  \& G.~{Beccari}, 179

\bibitem[{{Knigge} {et~al.}(2022){Knigge}, {Toonen}, \&
  {Boekholt}}]{knigge2022}
{Knigge}, C., {Toonen}, S., \& {Boekholt}, T.~C.~N. 2022, \mnras, 514, 1895

\bibitem[{{Kobulnicky} \& {Fryer}(2007)}]{kobulnicky2007}
{Kobulnicky}, H.~A. \& {Fryer}, C.~L. 2007, \apj, 670, 747

\bibitem[{{Kobulnicky} {et~al.}(2014){Kobulnicky}, {Kiminki}, {Lundquist},
  {Burke}, {Chapman}, {Keller}, {Lester}, {Rolen}, {Topel}, {Bhattacharjee},
  {Smullen}, {Vargas {\'A}lvarez}, {Runnoe}, {Dale}, \&
  {Brotherton}}]{kobulnicky2014}
{Kobulnicky}, H.~A., {Kiminki}, D.~C., {Lundquist}, M.~J., {et~al.} 2014,
  \apjs, 213, 34

\bibitem[{{Kochanek} {et~al.}(2014){Kochanek}, {Adams}, \&
  {Belczynski}}]{kochanek2014}
{Kochanek}, C.~S., {Adams}, S.~M., \& {Belczynski}, K. 2014, \mnras, 443, 1319

\bibitem[{{Kozai}(1962)}]{kozai1962}
{Kozai}, Y. 1962, \aj, 67, 591

\bibitem[{{Kummer} {et~al.}(2023){Kummer}, {Toonen}, \& {de
  Koter}}]{kummer2023}
{Kummer}, F., {Toonen}, S., \& {de Koter}, A. 2023, \aap, 678, A60

\bibitem[{{Langer}(2012)}]{langer2012}
{Langer}, N. 2012, \araa, 50, 107

\bibitem[{{Laplace} {et~al.}(2021){Laplace}, {Justham}, {Renzo}, {G{\"o}tberg},
  {Farmer}, {Vartanyan}, \& {de Mink}}]{laplace2021}
{Laplace}, E., {Justham}, S., {Renzo}, M., {et~al.} 2021, \aap, 656, A58

\bibitem[{{Leitherer} \& {Chavarria-K.}(1987)}]{leitherer1987}
{Leitherer}, C. \& {Chavarria-K.}, C. 1987, \aap, 175, 208

\bibitem[{{Li} {et~al.}(2025){Li}, {Lu}, {L{\"u}}, {Zhu}, {Liu}, \&
  {Yu}}]{li2025}
{Li}, Z., {Lu}, X., {L{\"u}}, G., {et~al.} 2025, \apjl, 979, L37

\bibitem[{{Li} {et~al.}(2024){Li}, {Zhu}, {Lu}, {L{\"u}}, {Li}, {Liu}, {Guo},
  \& {Yu}}]{li2024}
{Li}, Z., {Zhu}, C., {Lu}, X., {et~al.} 2024, \apjl, 975, L8

\bibitem[{{Lidov}(1962)}]{lidov1962}
{Lidov}, M.~L. 1962, \planss, 9, 719

\bibitem[{{Lombardi} {et~al.}(2002){Lombardi}, {Warren}, {Rasio}, {Sills}, \&
  {Warren}}]{lombardi2002}
{Lombardi}, Jr., J.~C., {Warren}, J.~S., {Rasio}, F.~A., {Sills}, A., \&
  {Warren}, A.~R. 2002, \apj, 568, 939

\bibitem[{{Mahy} {et~al.}(2020){Mahy}, {Sana}, {Abdul-Masih}, {Almeida},
  {Langer}, {Shenar}, {de Koter}, {de Mink}, {de Wit}, {Grin}, {Evans},
  {Moffat}, {Schneider}, {Barb{\'a}}, {Clark}, {Crowther}, {Gr{\"a}fener},
  {Lennon}, {Tramper}, \& {Vink}}]{mahy2020}
{Mahy}, L., {Sana}, H., {Abdul-Masih}, M., {et~al.} 2020, \aap, 634, A118

\bibitem[{{Mandel}(2021)}]{mandel2021}
{Mandel}, I. 2021, Research Notes of the American Astronomical Society, 5, 223

\bibitem[{{Mardling} \& {Aarseth}(1999)}]{mardling1999}
{Mardling}, R. \& {Aarseth}, S. 1999, in NATO Advanced Study Institute (ASI)
  Series C, Vol. 522, The Dynamics of Small Bodies in the Solar System, A Major
  Key to Solar System Studies, ed. B.~A. {Steves} \& A.~E. {Roy}, 385

\bibitem[{{Mardling} \& {Aarseth}(2001)}]{mardling2001}
{Mardling}, R.~A. \& {Aarseth}, S.~J. 2001, \mnras, 321, 398

\bibitem[{{McCrea}(1964)}]{mccrea1964}
{McCrea}, W.~H. 1964, \mnras, 128, 147

\bibitem[{{Menon} {et~al.}(2021){Menon}, {Langer}, {de Mink}, {Justham}, {Sen},
  {Sz{\'e}csi}, {de Koter}, {Abdul-Masih}, {Sana}, {Mahy}, \&
  {Marchant}}]{menon2021}
{Menon}, A., {Langer}, N., {de Mink}, S.~E., {et~al.} 2021, \mnras, 507, 5013

\bibitem[{{Menon} {et~al.}(2024){Menon}, {Pawlak}, {Lennon}, {Sen}, \&
  {Langer}}]{menon2024}
{Menon}, A., {Pawlak}, M., {Lennon}, D.~J., {Sen}, K., \& {Langer}, N. 2024,
  arXiv e-prints, arXiv:2410.16427

\bibitem[{{Michaely} \& {Perets}(2019)}]{michaely2019}
{Michaely}, E. \& {Perets}, H.~B. 2019, \mnras, 484, 4711

\bibitem[{{Moe} \& {Di Stefano}(2017)}]{moe2017}
{Moe}, M. \& {Di Stefano}, R. 2017, \apjs, 230, 15

\bibitem[{{Naoz}(2016)}]{naoz2016}
{Naoz}, S. 2016, \araa, 54, 441

\bibitem[{{Naoz} \& {Fabrycky}(2014)}]{naoz2014}
{Naoz}, S. \& {Fabrycky}, D.~C. 2014, \apj, 793, 137

\bibitem[{{Naoz} {et~al.}(2016){Naoz}, {Fragos}, {Geller}, {Stephan}, \&
  {Rasio}}]{naoz_xray2016}
{Naoz}, S., {Fragos}, T., {Geller}, A., {Stephan}, A.~P., \& {Rasio}, F.~A.
  2016, \apjl, 822, L24

\bibitem[{{Nelemans} {et~al.}(2000){Nelemans}, {Verbunt}, {Yungelson}, \&
  {Portegies Zwart}}]{Nelemans2000}
{Nelemans}, G., {Verbunt}, F., {Yungelson}, L.~R., \& {Portegies Zwart}, S.~F.
  2000, \aap, 360, 1011

\bibitem[{{Neumann} {et~al.}(2023){Neumann}, {Avakyan}, {Doroshenko}, \&
  {Santangelo}}]{neumann2023}
{Neumann}, M., {Avakyan}, A., {Doroshenko}, V., \& {Santangelo}, A. 2023, \aap,
  677, A134

\bibitem[{{Nieuwenhuijzen} \& {de Jager}(1990)}]{nieuwenhuijzen1990}
{Nieuwenhuijzen}, H. \& {de Jager}, C. 1990, \aap, 231, 134

\bibitem[{{Offner} {et~al.}(2023){Offner}, {Moe}, {Kratter}, {Sadavoy},
  {Jensen}, \& {Tobin}}]{offner2023}
{Offner}, S.~S.~R., {Moe}, M., {Kratter}, K.~M., {et~al.} 2023, in Astronomical
  Society of the Pacific Conference Series, Vol. 534, Protostars and Planets
  VII, ed. S.~{Inutsuka}, Y.~{Aikawa}, T.~{Muto}, K.~{Tomida}, \& M.~{Tamura},
  275

\bibitem[{{Pastorello} {et~al.}(2019){Pastorello}, {Mason}, {Taubenberger},
  {Fraser}, {Cortini}, {Tomasella}, {Botticella}, {Elias-Rosa}, {Kotak},
  {Smartt}, {Benetti}, {Cappellaro}, {Turatto}, {Tartaglia}, {Djorgovski},
  {Drake}, {Berton}, {Briganti}, {Brimacombe}, {Bufano}, {Cai}, {Chen},
  {Christensen}, {Ciabattari}, {Congiu}, {Dimai}, {Inserra}, {Kankare},
  {Magill}, {Maguire}, {Martinelli}, {Morales-Garoffolo}, {Ochner}, {Pignata},
  {Reguitti}, {Sollerman}, {Spiro}, {Terreran}, \& {Wright}}]{pastorello2019}
{Pastorello}, A., {Mason}, E., {Taubenberger}, S., {et~al.} 2019, \aap, 630,
  A75

\bibitem[{{Patton} {et~al.}(2025){Patton}, {Pinsonneault}, \&
  {Thompson}}]{patton2025}
{Patton}, R.~A., {Pinsonneault}, M.~H., \& {Thompson}, T.~A. 2025, \apj, 987,
  212

\bibitem[{{Pejcha}(2014)}]{pejcha2014}
{Pejcha}, O. 2014, \apj, 788, 22

\bibitem[{{Perets} \& {Fabrycky}(2009)}]{perets2009}
{Perets}, H.~B. \& {Fabrycky}, D.~C. 2009, \apj, 697, 1048

\bibitem[{{Peters}(1964)}]{peters1964}
{Peters}, P.~C. 1964, PhD thesis, California Institute of Technology

\bibitem[{{Pfahl} {et~al.}(2002){Pfahl}, {Rappaport}, {Podsiadlowski}, \&
  {Spruit}}]{pfahl2002}
{Pfahl}, E., {Rappaport}, S., {Podsiadlowski}, P., \& {Spruit}, H. 2002, \apj,
  574, 364

\bibitem[{{Podsiadlowski} {et~al.}(2004){Podsiadlowski}, {Langer},
  {Poelarends}, {Rappaport}, {Heger}, \& {Pfahl}}]{podsiadlowski2004}
{Podsiadlowski}, P., {Langer}, N., {Poelarends}, A.~J.~T., {et~al.} 2004, \apj,
  612, 1044

\bibitem[{{Podsiadlowski} {et~al.}(2003){Podsiadlowski}, {Rappaport}, \&
  {Han}}]{podsiadlowski2003}
{Podsiadlowski}, P., {Rappaport}, S., \& {Han}, Z. 2003, \mnras, 341, 385

\bibitem[{{Pols} \& {Marinus}(1994)}]{pols1994}
{Pols}, O.~R. \& {Marinus}, M. 1994, \aap, 288, 475

\bibitem[{{Pols} {et~al.}(1998){Pols}, {Schr{\"o}der}, {Hurley}, {Tout}, \&
  {Eggleton}}]{pols1998}
{Pols}, O.~R., {Schr{\"o}der}, K.-P., {Hurley}, J.~R., {Tout}, C.~A., \&
  {Eggleton}, P.~P. 1998, \mnras, 298, 525

\bibitem[{{Portegies Zwart} \& {Verbunt}(1996)}]{portegies1996}
{Portegies Zwart}, S.~F. \& {Verbunt}, F. 1996, \aap, 309, 179

\bibitem[{{Preece} {et~al.}(2024){Preece}, {Vigna-G{\'o}mez}, {Rajamuthukumar},
  {Vynatheya}, \& {Klencki}}]{preece2024}
{Preece}, H.~P., {Vigna-G{\'o}mez}, A., {Rajamuthukumar}, A.~S., {Vynatheya},
  P., \& {Klencki}, J. 2024, arXiv e-prints, arXiv:2412.14022

\bibitem[{{Reg{\'a}ly} {et~al.}(2025){Reg{\'a}ly}, {Fr{\"o}hlich}, \&
  {Vink{\'o}}}]{regaly2025}
{Reg{\'a}ly}, Z., {Fr{\"o}hlich}, V., \& {Vink{\'o}}, J. 2025, \apjl, 988, L7

\bibitem[{{Reimers}(1975)}]{reimers1975}
{Reimers}, D. 1975, in Problems in stellar atmospheres and envelopes., ed.
  B.~{Baschek}, W.~H. {Kegel}, \& G.~{Traving}, 229--256

\bibitem[{{Sana} {et~al.}(2012){Sana}, {de Mink}, {de Koter}, {Langer},
  {Evans}, {Gieles}, {Gosset}, {Izzard}, {Le Bouquin}, \&
  {Schneider}}]{sana2012}
{Sana}, H., {de Mink}, S.~E., {de Koter}, A., {et~al.} 2012, Science, 337, 444

\bibitem[{{Sana} \& {Vrancken}(2026)}]{sana2025}
{Sana}, H. \& {Vrancken}, J. 2026, in Encyclopedia of Astrophysics, Vol.~2,
  106--118

\bibitem[{{Schneider} {et~al.}(2019){Schneider}, {Ohlmann}, {Podsiadlowski},
  {R{\"o}pke}, {Balbus}, {Pakmor}, \& {Springel}}]{schneider2019}
{Schneider}, F. R.~N., {Ohlmann}, S.~T., {Podsiadlowski}, P., {et~al.} 2019,
  \nat, 574, 211

\bibitem[{{Schneider} {et~al.}(2016){Schneider}, {Podsiadlowski}, {Langer},
  {Castro}, \& {Fossati}}]{schneider2016}
{Schneider}, F.~R.~N., {Podsiadlowski}, P., {Langer}, N., {Castro}, N., \&
  {Fossati}, L. 2016, \mnras, 457, 2355

\bibitem[{{Schneider} {et~al.}(2024){Schneider}, {Podsiadlowski}, \&
  {Laplace}}]{schneider2024}
{Schneider}, F.~R.~N., {Podsiadlowski}, P., \& {Laplace}, E. 2024, \aap, 686,
  A45

\bibitem[{{Schneider} {et~al.}(2021){Schneider}, {Podsiadlowski}, \&
  {M{\"u}ller}}]{schneider2021}
{Schneider}, F.~R.~N., {Podsiadlowski}, P., \& {M{\"u}ller}, B. 2021, \aap,
  645, A5

\bibitem[{{Sch{\"o}ller} {et~al.}(2017){Sch{\"o}ller}, {Hubrig}, {Fossati},
  {Carroll}, {Briquet}, {Oskinova}, {J{\"a}rvinen}, {Ilyin}, {Castro}, {Morel},
  {Langer}, {Przybilla}, {Nieva}, {Kholtygin}, {Sana}, {Herrero}, {Barb{\'a}},
  {de Koter}, \& {BOB Collaboration}}]{scholler2017}
{Sch{\"o}ller}, M., {Hubrig}, S., {Fossati}, L., {et~al.} 2017, \aap, 599, A66

\bibitem[{{Sciarini} {et~al.}(2025){Sciarini}, {Ekstr{\"o}m}, {Kummer},
  {Rieder}, {Bruenech}, {Toonen}, \& {Farrell}}]{sciarini2025}
{Sciarini}, L., {Ekstr{\"o}m}, S., {Kummer}, F., {et~al.} 2025, \aap, 698, A240

\bibitem[{{Sepinsky} {et~al.}(2007){Sepinsky}, {Willems}, {Kalogera}, \&
  {Rasio}}]{sepinsky2007}
{Sepinsky}, J.~F., {Willems}, B., {Kalogera}, V., \& {Rasio}, F.~A. 2007, \apj,
  667, 1170

\bibitem[{{Sepinsky} {et~al.}(2009){Sepinsky}, {Willems}, {Kalogera}, \&
  {Rasio}}]{spinsky2009}
{Sepinsky}, J.~F., {Willems}, B., {Kalogera}, V., \& {Rasio}, F.~A. 2009, \apj,
  702, 1387

\bibitem[{{Shariat} {et~al.}(2025{\natexlab{a}}){Shariat}, {El-Badry}, \&
  {Naoz}}]{shariat_gaia2025}
{Shariat}, C., {El-Badry}, K., \& {Naoz}, S. 2025{\natexlab{a}}, \pasp, 137,
  094201

\bibitem[{{Shariat} {et~al.}(2025{\natexlab{b}}){Shariat}, {El-Badry}, {Naoz},
  {Rodriguez}, \& {van Roestel}}]{shariat_cv2025}
{Shariat}, C., {El-Badry}, K., {Naoz}, S., {Rodriguez}, A.~C., \& {van
  Roestel}, J. 2025{\natexlab{b}}, \pasp, 137, 074201

\bibitem[{{Shariat} {et~al.}(2025{\natexlab{c}}){Shariat}, {Naoz}, {El-Badry},
  {Rocha}, {Kalogera}, {Stephan}, {Burdge}, \& {Angelo}}]{shariat_xray2025}
{Shariat}, C., {Naoz}, S., {El-Badry}, K., {et~al.} 2025{\natexlab{c}}, \apj,
  983, 115

\bibitem[{{Shariat} {et~al.}(2025{\natexlab{d}}){Shariat}, {Naoz}, {El-Badry},
  {Rodriguez}, {Hansen}, {Angelo}, \& {Stephan}}]{shariat2025}
{Shariat}, C., {Naoz}, S., {El-Badry}, K., {et~al.} 2025{\natexlab{d}}, \apj,
  978, 47

\bibitem[{{Shariat} {et~al.}(2023){Shariat}, {Naoz}, {Hansen}, {Angelo},
  {Michaely}, \& {Stephan}}]{shariat2023}
{Shariat}, C., {Naoz}, S., {Hansen}, B. M.~S., {et~al.} 2023, \apjl, 955, L14

\bibitem[{{Shenar} {et~al.}(2022){Shenar}, {Sana}, {Mahy}, {El-Badry},
  {Marchant}, {Langer}, {Hawcroft}, {Fabry}, {Sen}, {Almeida}, {Abdul-Masih},
  {Bodensteiner}, {Crowther}, {Gieles}, {Gromadzki}, {H{\'e}nault-Brunet},
  {Herrero}, {de Koter}, {Iwanek}, {Koz{\l}owski}, {Lennon}, {Ma{\'\i}z
  Apell{\'a}niz}, {Mr{\'o}z}, {Moffat}, {Picco}, {Pietrukowicz}, {Poleski},
  {Rybicki}, {Schneider}, {Skowron}, {Skowron}, {Soszy{\'n}ski},
  {Szyma{\'n}ski}, {Toonen}, {Udalski}, {Ulaczyk}, {Vink}, \&
  {Wrona}}]{shenar2022}
{Shenar}, T., {Sana}, H., {Mahy}, L., {et~al.} 2022, Nature Astronomy, 6, 1085

\bibitem[{{Sills} {et~al.}(2005){Sills}, {Adams}, \& {Davies}}]{sills2005}
{Sills}, A., {Adams}, T., \& {Davies}, M.~B. 2005, \mnras, 358, 716

\bibitem[{{Sills} \& {Bailyn}(1999)}]{sills1999}
{Sills}, A. \& {Bailyn}, C.~D. 1999, \apj, 513, 428

\bibitem[{{Sills} {et~al.}(1997){Sills}, {Lombardi}, {Bailyn}, {Demarque},
  {Rasio}, \& {Shapiro}}]{sills1997}
{Sills}, A., {Lombardi}, Jr., J.~C., {Bailyn}, C.~D., {et~al.} 1997, \apj, 487,
  290

\bibitem[{{Silsbee} \& {Tremaine}(2017)}]{silsbee2017S}
{Silsbee}, K. \& {Tremaine}, S. 2017, \apj, 836, 39

\bibitem[{{Smith} {et~al.}(2018){Smith}, {Andrews}, {Rest}, {Bianco}, {Prieto},
  {Matheson}, {James}, {Smith}, {Strampelli}, \& {Zenteno}}]{smith2018}
{Smith}, N., {Andrews}, J.~E., {Rest}, A., {et~al.} 2018, \mnras, 480, 1466

\bibitem[{{Smith} {et~al.}(2016){Smith}, {Andrews}, {Van Dyk}, {Mauerhan},
  {Kasliwal}, {Bond}, {Filippenko}, {Clubb}, {Graham}, {Perley}, {Jencson},
  {Bally}, {Ubeda}, \& {Sabbi}}]{smith2016}
{Smith}, N., {Andrews}, J.~E., {Van Dyk}, S.~D., {et~al.} 2016, \mnras, 458,
  950

\bibitem[{{Soker} \& {Tylenda}(2003)}]{soker2003}
{Soker}, N. \& {Tylenda}, R. 2003, \apjl, 582, L105

\bibitem[{{Stegmann} {et~al.}(2022{\natexlab{a}}){Stegmann}, {Antonini}, \&
  {Moe}}]{stegmann2022a}
{Stegmann}, J., {Antonini}, F., \& {Moe}, M. 2022{\natexlab{a}}, \mnras, 516,
  1406

\bibitem[{{Stegmann} {et~al.}(2022{\natexlab{b}}){Stegmann}, {Antonini},
  {Schneider}, {Tiwari}, \& {Chattopadhyay}}]{stegmann2022b}
{Stegmann}, J., {Antonini}, F., {Schneider}, F. R.~N., {Tiwari}, V., \&
  {Chattopadhyay}, D. 2022{\natexlab{b}}, \prd, 106, 023014

\bibitem[{{Stegmann} \& {Klencki}(2025)}]{stegmann2025}
{Stegmann}, J. \& {Klencki}, J. 2025, arXiv e-prints, arXiv:2506.09121

\bibitem[{{Stegmann} {et~al.}(2024){Stegmann}, {Vigna-G{\'o}mez}, {Rantala},
  {Wagg}, {Zwick}, {Renzo}, {van Son}, {de Mink}, \& {White}}]{stegmann2024}
{Stegmann}, J., {Vigna-G{\'o}mez}, A., {Rantala}, A., {et~al.} 2024, \apjl,
  972, L19

\bibitem[{{Suzuki} {et~al.}(2007){Suzuki}, {Nakasato}, {Baumgardt},
  {Ibukiyama}, {Makino}, \& {Ebisuzaki}}]{suzuki2007}
{Suzuki}, T.~K., {Nakasato}, N., {Baumgardt}, H., {et~al.} 2007, \apj, 668, 435

\bibitem[{{Tauris} \& {van den Heuvel}(2023)}]{tauris2023}
{Tauris}, T.~M. \& {van den Heuvel}, E. P.~J. 2023, {Physics of Binary Star
  Evolution. From Stars to X-ray Binaries and Gravitational Wave Sources}

\bibitem[{{Tejeda} \& {Toal{\'a}}(2025)}]{tejeda2025}
{Tejeda}, E. \& {Toal{\'a}}, J.~A. 2025, \apj, 980, 226

\bibitem[{{Temmink} {et~al.}(2020){Temmink}, {Toonen}, {Zapartas}, {Justham},
  \& {G{\"a}nsicke}}]{temmink2020}
{Temmink}, K.~D., {Toonen}, S., {Zapartas}, E., {Justham}, S., \&
  {G{\"a}nsicke}, B.~T. 2020, \aap, 636, A31

\bibitem[{{Tokovinin}(2017)}]{tokovinin2017}
{Tokovinin}, A. 2017, \apj, 844, 103

\bibitem[{{Toonen} {et~al.}(2016){Toonen}, {Hamers}, \& {Portegies
  Zwart}}]{toonen2016}
{Toonen}, S., {Hamers}, A., \& {Portegies Zwart}, S. 2016, Computational
  Astrophysics and Cosmology, 3, 6

\bibitem[{{Toonen} {et~al.}(2017){Toonen}, {Hollands}, {G{\"a}nsicke}, \&
  {Boekholt}}]{toonen2017}
{Toonen}, S., {Hollands}, M., {G{\"a}nsicke}, B.~T., \& {Boekholt}, T. 2017,
  \aap, 602, A16

\bibitem[{{Toonen} \& {Nelemans}(2013)}]{toonen2013}
{Toonen}, S. \& {Nelemans}, G. 2013, \aap, 557, A87

\bibitem[{{Toonen} {et~al.}(2012){Toonen}, {Nelemans}, \& {Portegies
  Zwart}}]{toonen2012}
{Toonen}, S., {Nelemans}, G., \& {Portegies Zwart}, S. 2012, \aap, 546, A70

\bibitem[{{Toonen} {et~al.}(2020){Toonen}, {Portegies Zwart}, {Hamers}, \&
  {Bandopadhyay}}]{toonen2020}
{Toonen}, S., {Portegies Zwart}, S., {Hamers}, A.~S., \& {Bandopadhyay}, D.
  2020, \aap, 640, A16

\bibitem[{{Tylenda} {et~al.}(2011){Tylenda}, {Hajduk}, {Kami{\'n}ski},
  {Udalski}, {Soszy{\'n}ski}, {Szyma{\'n}ski}, {Kubiak}, {Pietrzy{\'n}ski},
  {Poleski}, {Wyrzykowski}, \& {Ulaczyk}}]{tylenda2011}
{Tylenda}, R., {Hajduk}, M., {Kami{\'n}ski}, T., {et~al.} 2011, \aap, 528, A114

\bibitem[{{Tylenda} \& {Soker}(2006)}]{tylenda2006}
{Tylenda}, R. \& {Soker}, N. 2006, \aap, 451, 223

\bibitem[{{van Son} {et~al.}(2025){van Son}, {Roy}, {Mandel}, {Farr}, {Lam},
  {Merritt}, {Broekgaarden}, {Sander}, \& {Andrews}}]{vanson2025}
{van Son}, L.~A.~C., {Roy}, S.~K., {Mandel}, I., {et~al.} 2025, \apj, 979, 209

\bibitem[{{Vassiliadis} \& {Wood}(1993)}]{vassiliadis1993}
{Vassiliadis}, E. \& {Wood}, P.~R. 1993, \apj, 413, 641

\bibitem[{{Verbunt} {et~al.}(2017){Verbunt}, {Igoshev}, \&
  {Cator}}]{verbunt2017}
{Verbunt}, F., {Igoshev}, A., \& {Cator}, E. 2017, \aap, 608, A57

\bibitem[{{Vigna-G{\'o}mez} {et~al.}(2025){Vigna-G{\'o}mez}, {Grishin},
  {Stegmann}, {Olejak}, {Popa}, {Liu}, {Rajamuthukumar}, {van Son}, {Bobrick},
  \& {Dorozsmai}}]{vigna-gomez2025}
{Vigna-G{\'o}mez}, A., {Grishin}, E., {Stegmann}, J., {et~al.} 2025, \aap, 699,
  A272

\bibitem[{{Vigna-G{\'o}mez} {et~al.}(2024){Vigna-G{\'o}mez}, {Willcox},
  {Tamborra}, {Mandel}, {Renzo}, {Wagg}, {Janka}, {Kresse}, {Bodensteiner},
  {Shenar}, \& {Tauris}}]{vigna-gomez2024}
{Vigna-G{\'o}mez}, A., {Willcox}, R., {Tamborra}, I., {et~al.} 2024, \prl, 132,
  191403

\bibitem[{{Vink} {et~al.}(2001){Vink}, {de Koter}, \& {Lamers}}]{vink2001}
{Vink}, J.~S., {de Koter}, A., \& {Lamers}, H.~J.~G.~L.~M. 2001, \aap, 369, 574

\bibitem[{{Vink} \& {Sander}(2021)}]{vink2021}
{Vink}, J.~S. \& {Sander}, A. A.~C. 2021, \mnras, 504, 2051

\bibitem[{{von Zeipel}(1910)}]{vonzeipel1910}
{von Zeipel}, H. 1910, Astronomische Nachrichten, 183, 345

\bibitem[{{Wade} \& {MiMeS Collaboration}(2015)}]{wade2015}
{Wade}, G.~A. \& {MiMeS Collaboration}. 2015, in Astronomical Society of the
  Pacific Conference Series, Vol. 494, Physics and Evolution of Magnetic and
  Related Stars, ed. Y.~Y. {Balega}, I.~I. {Romanyuk}, \& D.~O. {Kudryavtsev},
  30

\bibitem[{{Webbink}(1984)}]{webbink1984}
{Webbink}, R.~F. 1984, \apj, 277, 355

\bibitem[{{Zahn} {et~al.}(1997){Zahn}, {Talon}, \& {Matias}}]{zahn1997}
{Zahn}, J.~P., {Talon}, S., \& {Matias}, J. 1997, \aap, 322, 320

\bibitem[{{Zapartas} {et~al.}(2017){Zapartas}, {de Mink}, {Izzard}, {Yoon},
  {Badenes}, {G{\"o}tberg}, {de Koter}, {Neijssel}, {Renzo}, {Schootemeijer},
  \& {Shrotriya}}]{zapartas2017}
{Zapartas}, E., {de Mink}, S.~E., {Izzard}, R.~G., {et~al.} 2017, \aap, 601,
  A29

\bibitem[{{Zorotovic} {et~al.}(2010){Zorotovic}, {Schreiber}, {G{\"a}nsicke},
  \& {Nebot G{\'o}mez-Mor{\'a}n}}]{zorotovic2010}
{Zorotovic}, M., {Schreiber}, M.~R., {G{\"a}nsicke}, B.~T., \& {Nebot
  G{\'o}mez-Mor{\'a}n}, A. 2010, \aap, 520, A86

\end{thebibliography}

\begin{appendix}
\onecolumn
\section{Evolutionary channel rates}
\begin{table}[h!]
\caption{Percentage of stellar mergers and post-merger interactions within our simulated populations of massive hierarchical triples for different model variations.}           
\label{table:merger_fraction}  
\centering                         
\resizebox{\textwidth}{!}{\begin{tabular}{l c c c c c}        
\hline\hline  
Model & Mergers (\%) & Non-interacting (\%) & MT to merger companion (\%) & MT to merger remnant (\%) & Merger Rate ($10^{-3} \, \rm{yr}^{-1}$) \\
\hline\hline
\textbf{Fiducial} & 28.7 & 82.7 & 13.7 & 3.7 & 1.9 \\
\textbf{Conservative MT} & 20.4 & 78.2 & 18.1 & 3.7 & 1.3 \\
\textbf{Inefficient CE \boldmath ($\alpha\lambda=0.25$)} & 31.8 & 84.3 & 12.4 & 3.3 & 2.1 \\
\textbf{\boldmath $\tau_{\rm{CE}}=10^4\,\rm{yr}$} & 28.7 & 82.6 & 13.7 & 3.7 & 1.8 \\
\textbf{Non-conservative merger} & 28.4 & 83.2 & 12.9 & 3.9 & 1.8 \\
\textbf{\boldmath $q_{\rm{out}}=m_3/m_1$} & 31.6 & 82.0 & 16.5 & 1.5 & 2.2 \\
\hline\hline                                 
\end{tabular}}
\end{table}
\end{appendix}

\end{document}